\documentclass{aa}
\usepackage{graphicx}
\usepackage{subcaption}
\usepackage{float}
\usepackage{txfonts}
\usepackage{amssymb}
\usepackage{amsmath}
\usepackage{epsfig}
\usepackage{epstopdf}
\usepackage{times}
\usepackage{url}
\usepackage{color}
\usepackage{placeins}
\usepackage{hyperref}
\usepackage{natbib}
\usepackage{xcolor}

\begin{document} 

\title{Eruptions and Flaring Activity in Emerging Quadrupolar Regions}
\subtitle{}
\author{P. Syntelis \inst{1}
        \and
      E. J. Lee \inst{1}
        \and
      C. W. Fairbairn \inst{1}
        \and
      V. Archontis \inst{1} 
        \and        
      A. W. Hood \inst{1}
       }

\institute{  School of Mathematics and Statistics,
            University of St Andrews, 
            St Andrews, Fife, KY16 9SS, UK \\
            \email{ps84@st-andrews.ac.uk}
         }

\date{\today}

\abstract
{
    Solar observations suggest that some of the most dynamic active regions are associated with complex photospheric magnetic configurations such as quadrupolar regions, and especially ones having a $\delta$-spot configuration and a strong Polarity Inversion Line (PIL). 
}
{
    We study the formation and eruption of magnetic flux ropes in quadrupolar regions.
}
{
    We perform 3D MHD simulations of the partial emergence of a highly twisted flux tube from the solar interior into a non-magnetized stratified atmosphere. We introduce a density deficit at two places along the length of the subphotospheric flux tube to emerge two $\Omega$-shaped loops, forming a quadrupolar region. 
}
{
    At the photosphere, the emerging flux forms two initially separated bipoles, that later come in contact creating a $\delta$-spot central region. Above the two bipoles, two magnetic lobes expand and interact through a series of current sheets at the interface between them. Two recurrent confined eruptions are produced. 
    In both cases, reconnection between sheared low-lying field lines forms a flux rope. 
    Reconnection between the two lobes higher in the atmosphere forms field lines that retract down and push against the flux rope, creating a current sheet between them.
    It also forms field lines that create a third magnetic lobe between the two emerged ones, that later acts as a strapping field. 
    The flux rope eruptions are triggered when the reconnection between the flux ropes and the field above them becomes efficient enough to remove the tension of the overlying field. 
    These reconnection events occur internally in the quadrupolar system, as the atmosphere is non-magnetized.
    The flux rope of the first, weaker, eruption almost fully reconnects with the overlying field. 
    The flux rope of the second, more energetic, eruption is confined by the overlying strapping field.
    During the second eruption, the flux rope is enhanced in size, flux and twist, similar to confined-flare-to-flux-rope observations.
    Proxies of the emission reveal the two erupting filaments channels. A flare arcade is formed only in the second eruption due to the longer lasting and more efficient reconnection at the current sheet below the flux rope.
}
{}

\maketitle


\section{Introduction}


Flux emergence is a key process, which transfers magnetic energy from the solar interior, through the photosphere, up into the solar corona. 
It has long been posited that the buoyant rise of magnetic flux tubes could be responsible for the production of active regions \citep{Parker_1955}. 
Active regions are highly associated with a plethora of solar phenomena, such as flares and coronal mass ejections (CMEs), whereby stored free magnetic energy is released in an explosive manner \citep[e.g.][]{Priest_Forbes_2002}. The amount of stored and released energy can range significantly depending on the specifics of the active regions, leading to more or less energetic eruptions or flares.

The most intense solar phenomena are commonly associated with the presence of a strong, i.e. high-gradient, PIL \citep[e.g.][]{Schrijver_2009}. 
It is no surprise, therefore, that large active regions with strong PILs, such as NOAA 11429 \citep[e.g.][]{Liu_2014, Wang_etal2014, Chintzoglou_etal2015, Syntelis_etal2016,Patsourakos_etal2016,Zheng_etal2017,Yang_etal2018,Baker_etal2019,Zhou_etal2019} and the quadrapolar NOAA 11158 \citep[e.g.][]{Schrijver_etal2011b,Sun_etal2012,Wang_etal2012,Liu_etal2012,Vemareddy_etal2012,Dalmasse_etal2013,Tziotziou_etal2013,Janvier_etal2014,Kazachenko_etal2015,Zhang_etal2016} have been extensively studied due to the strong activity occurring during their long lifetimes.
Prominent examples where strong PILs are commonly found are the so-called $\delta$-spots \citep[][]{Kunzel_1960}. These are regions where the two opposite polarities are so ``compact'', that in white light observations the umbra of the two polarities share a common penumbra. $\delta$-spots are highly associated with intense solar activity and strong X-class flares \citep[e.g.][]{Zirin_etal1987,Schrijver_2007,Guo_etal2014}.
Strong PILs and $\delta$-spots can develop in a variety of different photospheric magnetic configurations. For instance, they can be formed between two main opposite polarities in a bipolar region, between the inner-most polarities of a quadrupolar active regions, between a main polarity and a parasitic polarity, between polarities of very compex shape, or even between polarities of separate active regions \citep[e.g.][]{Linton_etal1999,Fan_1999,Takasao_etal2015,Fang_Fan_2015,Toriumi_etal2017,Toriumi_etal2017b,Knizhnik_etal2018,Chintzoglou_etal2019}. 

The strong shearing along such strong PILs, the rotation of the polarities and/or the direct emergence of non-potential field, injects shear into the solar atmosphere, building up free magnetic energy, part of which accumulates along the PIL. 
This free energy is usually stored either in a sheared magnetic arcade \citep[e.g.][]{Antiochos_etal1999,Lynch_etal2008} or a magnetic flux rope \citep[e.g.][]{vanBallegooijen_etal1989,Torok_Kliem_2005}.
When such configurations are destabilized, they release part of their free energy in the form of a flare arcade and an accelerating magnetic flux rope (either a pre-existing or formed on-the-fly) whose flux and twist will be enhanced during the eruption \citep[e.g.][]{Inoue_etal2018,Syntelis_etal2019}.
However, such events are not always associated with the runaway ejection of plasma such as a CME. 
Depending on the specifics of the eruption, the erupting field can move upwards without being stopped (ejective eruption, eruptive flare) \citep[e.g.][]{Qiu_etal2005,Vourlidas_etal2012,Mitra_etal2018,Georgoulis_etal2019}, or it can be decelerated and become confined by an overlying strapping field (confined eruption and flare) \citep[e.g.][]{Moore_etal2001, Patsourakos_etal2013,Liu_etal2018}.

In order for an ejective eruption to occur, it is required that either the strength of the overlying field is decreasing rapidly \citep{Kliem_etal2006}, or that the overlying magnetic field is sufficiently removed. The latter, can occur in two ways.
One way is through external reconnection (e.g. breakout), whereby the strapping field reconnects with an external ambient atmospheric field. If the orientation of the external field favours reconnection, the strapping field can be removed leading to an ejective eruption. On the other hand, if the orientation does not favour reconnection, the strapping field will not be efficiently removed and its downwards tension will suppress the eruption \citep[e.g.][]{Archontis_etal2012,Leake_etal2013,Leake_etal2014}.
Another way that the tension of the overlying field can be removed is through internal reconnection. If an accelerating upwards magnetic flux rope stretches its strapping field enough, then, below the magnetic flux rope, the field lines of the strapping field can become anti-parallel and reconnect though a current sheet
\citep[tether-cutting reconnection, e.g.][]{Moore_etal1992}. 
The orientation of the field lines is such that this reconnection will be very efficient and will accelerate the flux rope further upwards, while removing the strapping field in a runaway manner \citep[e.g.][]{Moore_etal2001}. 

Numerical simulations focusing on the self-consistent coupling of the solar interior with the solar atmosphere have extensively studied the role of shearing, reconnection, and also ideal instabilities, in forming eruptions from bipolar regions with strong PILs \citep[e.g.][]{Manchester_etal2004, An_Magara_2013, Archontis_Torok2008,Archontis_etal2012, Leake_etal2013, Leake_etal2014,Syntelis_etal2017,Syntelis_etal2019,Archontis_etal2019}.
Such bipolar regions with compact PILs are usually formed when a highly twisted, but kink stable, single $\Omega$-loop flux tube partially emerges (i.e. its axis remain below/at the photosphere) from the solar interior into the solar atmosphere \citep[e.g.][]{fan_2001}.

Other configurations producing more complex regions with strong PILs have been also examined.
For instance, the emergence of a weakly- or non-twisted single $\Omega$-loop flux tubes can form a quadrupolar region, where the innermost polarities form a compact PIL \citep{Murray_etal2006, Archontis_etal2013,Syntelis_etal2015}. Such simulations can mimic observations of quadrupolar regions where the two bipoles emerge simultaneously, approach each other, and form a strong PIL in between them. 
Reconnection at the current sheet above these PILs can lead to the formation of post-emergence low-lying flux ropes.
In these examples the flux ropes remained stable.

Another approach to form quadrupolar regions is to assume a highly twisted subphotospheric flux tube that is buoyant at two locations along its length, leading to the emergence of two $\Omega$-loop segments. The segments partially emerge above the photosphere simultaneously, forming two bipoles. Again, the two innermost polarities collide forming a compact PILs similar to $\delta$-spots. It has been demonstrated that such configurations can build up shear and free energy in the corona, without however reporting eruptions and flares \citep{Fang_Fan_2015, Toriumi_etal2017b}.

Other numerical studies have shown that $\delta$-spots can also be formed when a kink unstable subphotospheric flux tube partially emerges into the atmosphere \citep[e.g.][]{Fan_1999,Linton_etal1999}. Depending on the amount of twist of the kink-unstable flux tube, either bipolar or quadrupolar regions with compact PILs can be formed \citep[e.g.][]{Knizhnik_etal2018}. The shearing injected in the atmosphere is mainly associated with the photospheric vorticity, and the polarities can be flux imbalanced.
In contrast, in the two $\Omega$-loops emergence of a kink stable flux tube described in the previous paragraphs, the shearing is mostly due to the horizontal motions of the polarities and the flux is mostly balanced \citep{Takasao_etal2015}. 
Configurations from kink unstable flux tubes can build up free energy in the atmosphere and produce sheared structures, however, eruptions or flare have not been studied yet.

\citet{Toriumi_etal2017b}, examined both the emergence of a kink unstable flux tube and a double $\Omega$-loop flux tube. They also examined the emergence of a smaller flux tube next to a larger flux tube to mimic a parasitic polarity emerging next to a developed active region, and the simultaneous emergence of two separate flux tubes next to each other, mimicking two nearby active regions. The most compact PILs were found in the kink unstable flux tubes, followed by the quadrupolar regions. All cases were able to build up free energy and sheared arcade fields, without however producing flares or eruptions.

Despite the significant progress in understanding the self-consistent formation of quadrupolar $\delta$-spot regions, eruptivity and flaring as not well understood. \cite{Lee_etal2015} investigated the emergence of a double $\Omega$-loop flux tube into a corona with an overlying field, giving rise to recurrent blowout jets driven by flux rope eruptions. However, the presence of a pre-existing coronal field significantly modifies the eruption dynamics. Thus, it is of interest to perform this experiment with a non-magnetized atmosphere and study the eruptions and flaring associated solely with the emerged quadrupolar region.
We thus aim to examine further the formation of flux ropes in quadrupolar regions and the triggering mechanism of the associated eruptions.

The paper is structured as follows. Section \ref{sec:model} describes the numerical setup used in the experiment. Section \ref{subsec:emergence} discusses the initial rise and emergence of the flux tube before sections \ref{subsec:first_eruption} and \ref{subsec:second_eruption} look in more detail at the mechanisms underpinning two eruptions. Section \ref{subsec:flare} discusses the flaring associated with these eruptions. All the findings are summarised and discussed in \ref{sec:discussion}.


\section{Numerical Setup}
\label{sec:model}
\subsection{MHD Equations}
In order to investigate flux emergence and the following dynamics, we employ a three-dimensional (3D) Lagrangian-Remap code. Lare3D, developed by \cite{Arber_etal2001}, solves the time-dependent, resistive and compressible MHD equations in a Cartesian geometry. The equations can be written in dimensionless form via the choice of normalisation constants which in this case are motivated by typical photospheric values: $\rho_{0} = 1.67\times 10^{-7}\,\mathrm{g}\,\mathrm{cm}^{-3}$, characteristic scale height $H_{0} = 180\,\mathrm{km}$ and the magnetic field strength $B_{0} = 300\,\mathrm{G}$. This in turn constrains the remaining normalisation constants such that pressure $P_{0} = 7.16\times 10^{3}\,\mathrm{erg}\,\mathrm{cm}^{-3}$, temperature $T_{0} = 623\,\mathrm{K}$, velocity $v_{0}=2.1\,\mathrm{km}\,\mathrm{s}^{-1}$ and time $t_{0} = 85.7\,\mathrm{s}$. The dimensionless equations become:
\begin{align}
    &\frac{\partial\rho}{\partial t} + \nabla\cdot(\rho\mathbf{v}) = 0,\\
    &\frac{\partial(\rho\mathbf{v})}{\partial t} = 
    -\nabla\cdot(\rho\mathbf{v}\mathbf{v})
    +(\nabla\times\mathbf{B})\times\mathbf{B}-\nabla P
     +\rho g+\nabla\cdot\mathbf{S}, \\
    &\frac{\partial(\rho\epsilon)}{\partial t} = 
    - \nabla\cdot(\rho\epsilon\mathbf{v})
    - P \nabla\cdot\mathbf{v}
    + Q_{\text{joule}} + Q_{\text{visc}}, \\
    &\frac{\partial{B}}{\partial t} = 
    \nabla\times(\mathbf{v}\times\mathbf{B})+\eta\nabla^2\mathbf{B}, \\
    &\epsilon = \frac{P}{(\gamma-1)\rho}.
\end{align}
Here, $\rho$, $\mathbf{v}$, $\mathbf{B}$, $P$ and $\epsilon$ are the density, velocity vector, magnetic field vector, gas pressure and specific internal energy  respectively. We assume an ideal gas with $\gamma = 5/3$  and a uniform gravitational field. $\mathbf{S}$ is the viscous stress tensor and $\eta$ is the resistivity resulting in joule dissipation of current. The resistivity is set to be a constant background value, $\eta = 0.01$, throughout the numerical domain. This high value ensures numerical stability and leads to suitable energy dissipation across the short length scales observed during the simulation. $Q_{\text{visc}}$ and $Q_{\text{joule}}$ are the viscous and ohmic heating respectively. 

\subsection{The Model}

\begin{figure}
    \centering
    \includegraphics[width=\columnwidth]{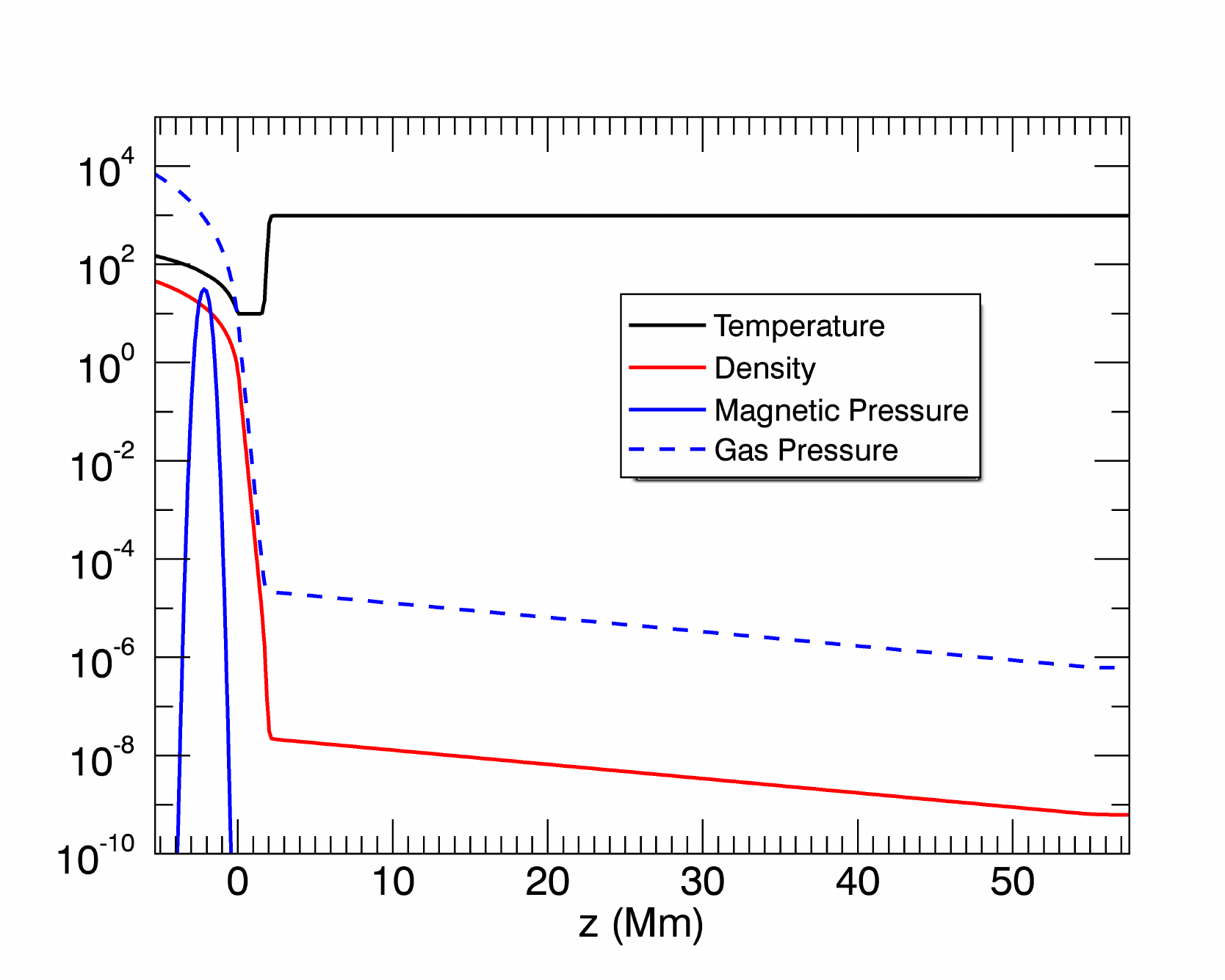}
    \caption{
        Initial stratification of the solar atmosphere in our numerical experiments, in dimensionless logarithmic scale. Temperature (black), density (red), magnetic pressure (solid blue) and gas pressure (dashed blue).
    }
    \label{fig:initial_conditions}
\end{figure}

The experimental setup is similar to \cite{Lee_etal2015}, with the key difference that we impose an unmagnetized corona. 
Figure \ref{fig:initial_conditions} shows the initial background stratification of the atmosphere. The atmosphere is in hydrostatic equilibrium. 
We implement an adiabatically stratified sub-photospheric region between $-5.4\,\mathrm{Mm}\le z < 0\,\mathrm{Mm}$ which is marginally stable to the convective instability. 
The photosphere/chromosphere is captured by an isothermal region between $0 \,\mathrm{Mm} \leq z \leq 1.9\,\mathrm{Mm}$. Above this lies the transition region where the temperature rises steeply between $1.9\,\mathrm{Mm} \leq z \leq 2.7\,\mathrm{Mm}$ and connects to the lower solar corona, which extends from $2.7\,\mathrm{Mm} \leq z \leq 57.6\,\mathrm{Mm}$.

A horizontal magnetic flux rope is placed within the solar interior at $z = -2.1\,\mathrm{Mm}$ oriented along the $y$-direction. The axial field strength falls off in a Gaussian manner away from the center of the flux tube, while the azimuthal field is such that the flux tube has constant twist:

\begin{align}
    &B_{y} = B_\mathrm{0} \exp(-r^2/R^2), \\
    &B_{\theta} = \alpha r B_{y}. 
\end{align}

The radius of the tube is taken to be $R = 450\,\mathrm{km}$ and $r$ is the radial distance from the axis. 
$B_{0}$ is the field strength along the tube axis and is set to be $2.4\,\mathrm{kG}$. The twist is characterized by a uniform value of $\alpha = 2.2\times10^{-3}\,\mathrm{km}^{-1}$ such that the tube is marginally stable to the kink instability. There is no pre-existing magnetic field in the corona therefore there is no interaction between the emerging field and any ambient background field.

In order to emerge a parts of the flux tube, a density deficit $\Delta\rho$ is imposed along the flux tube akin to the method used by \cite{fan_2001}. In this work, instead of emerging only the central part of the flux tube, we initiate the buoyant emergence in two locations along the length of the flux tube, via the following profile:
\begin{equation}
\Delta \rho = [p(r)/p_{b}(z)-1]\rho_{b}(z)\exp\left(-y^2/\lambda^2\right)\sin^2\left(2\pi y/\omega\right),
\end{equation}
where $p$ is the gas pressure within the tube, $p_{b}(z)$ is the pressure of the background atmosphere, and $\rho_{b}(z)$ is the density of the background atmosphere. $\lambda=3.6\,\mathrm{Mm}$ is the length of the buoyant part of the tube, and $\omega=31.5\,\mathrm{Mm}$ is half of the flux tube's length.

The simulation is carried out using a uniform numerical grid made up of $420^3$ grid points. This represents a physical domain of $[-31.5,31.5] \times [-31.5,31.5] \times [-5.4,57.6]\,\mathrm{Mm}$ in the horizontal direction perpendicular to the flux tube ($x$), along the flux tube axis ($y$) and the vertical direction ($z$), respectively. Periodic boundary conditions are used in the $y$ direction whilst the upper $z$, and both the $x$ boundaries, are implemented via open far-field Riemann characteristics, which allows plasma to flow out of the grid. The lower $z$ boundary is set to be a closed boundary.


\section{Results}
\label{sec:results}

\subsection{Initial Evolution and Emergence}
\label{subsec:emergence}

\begin{figure}
    \centering
    \includegraphics[width=\linewidth]{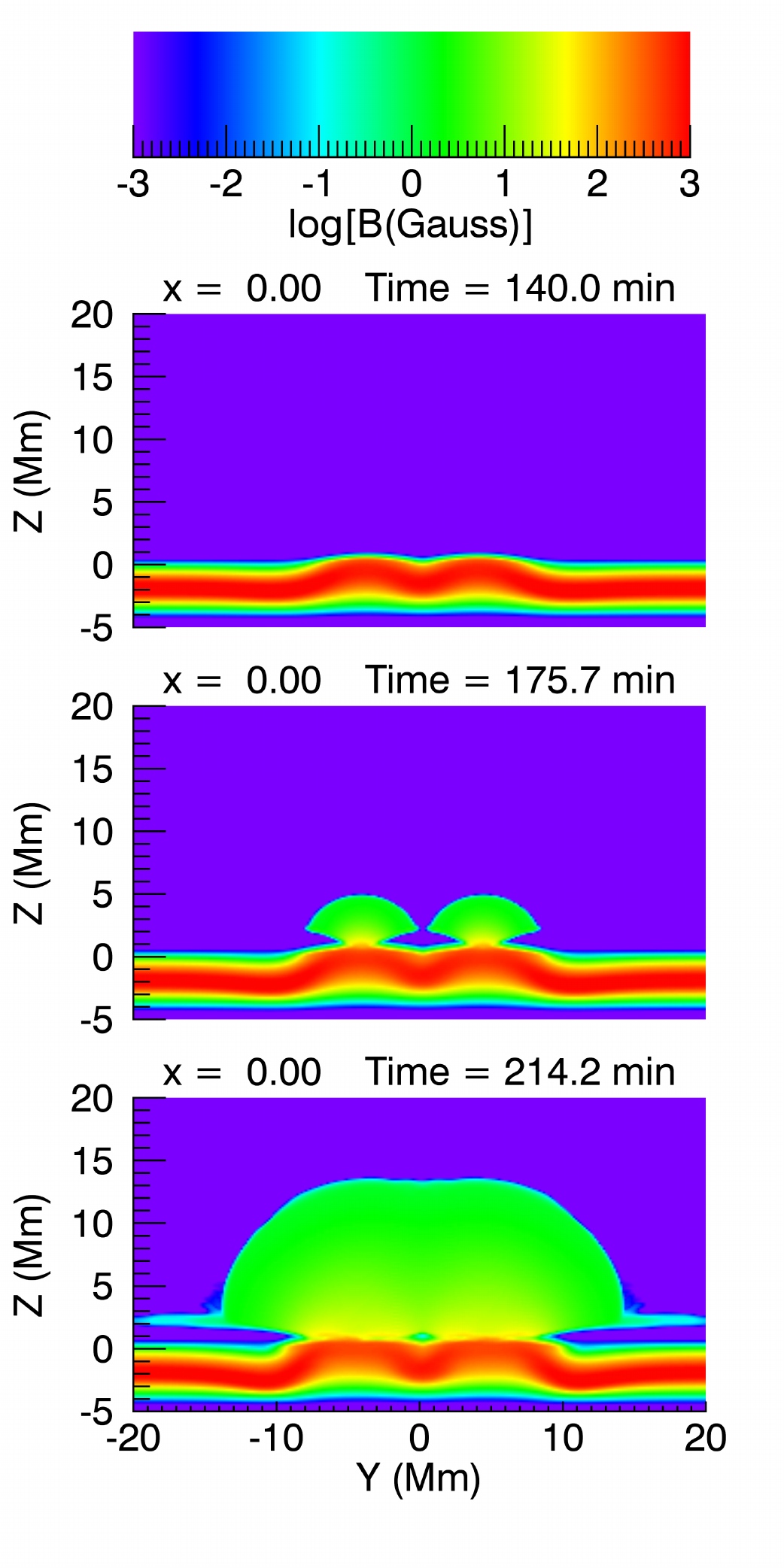}
    \caption{
        The absolute magnetic field strength plotted at the central $yz$ plane ($x = 0.0~\mathrm{Mm}$) intersecting the middle of the flux tube. 
        Upper panel shows the emerging segments of the flux tube becoming compressed below the photosphere ($t = 140.0\,\mathrm{min}$). 
        Middle panel shows the initial emergence of the field above the photosphere, leading to the formation of two magnetic lobes which extend into the corona ($t = 175.7\,\mathrm{min}$).
        Lower panel shows the later evolution of the two magnetic lobes, where they have further expanded, compressed against each other and eventually interacted ($t = 214.2\,\mathrm{min}$).
    }
    \label{fig:bcross}
\end{figure}

We will first briefly describe the overall evolution of our numerical experiment. The density deficit introduced to the flux tube means it is buoyantly unstable at two locations along its length. 
This causes two segments of the tube to rise, forming two $\Omega$-shaped loops (see magnetic field strength distribution, upper panel, Fig.~\ref{fig:bcross}). 
These two emerging loops continue to rise until the apex of each loop reach the photosphere ($z=0\,\mathrm{Mm}$). 
There, both loops compress and expand horizontally, similar to a single rising $\Omega$-loop \citep[e.g.][]{Archontis_etal2004}. 
This compression increases the apex magnetic field strength \citep[e.g.][]{Syntelis_etal2019a}. Eventually, around $t = 170.0\,\mathrm{min}$, the magnetic buoyancy instability is triggered \citep{Acheson1979,Archontis_etal2004}. 
After this, the apex field of both segments partially emerge above the photosphere and into the solar atmosphere forming two magnetic lobes (middle panel, Fig.~\ref{fig:bcross}). 

\begin{figure}
    \centering
    \includegraphics[width=\columnwidth]{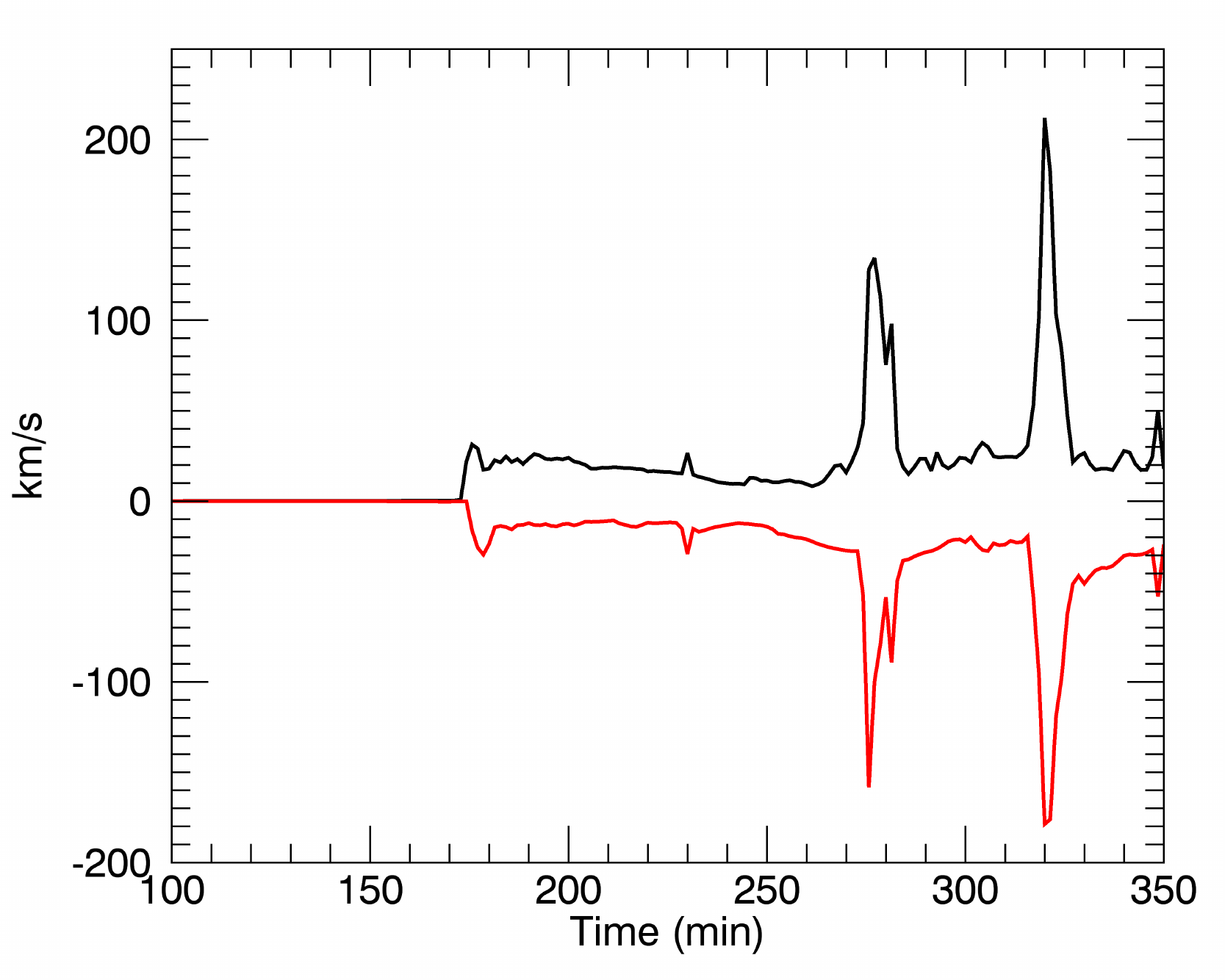}
    \caption{
        Temporal evolution of maximum (black) and minimum (red) $v_z$ in the corona. 
    }
    \label{fig:vz_time}
\end{figure}

As the flux continues to emerge, the two magnetic lobes expand both vertically and horizontally. The lobes eventually compress against each other, and later reconnect to form an envelope field which encloses the newly emerging flux (e.g. lower panel, Fig.~\ref{fig:bcross}). 
The interaction of the magnetic lobes triggers dynamical behaviour throughout the system. 
The maximum and minimum vertical velocities in the computational domain are plotted as a function of time in Fig.~\ref{fig:vz_time}, showing a series of plasma accelerations. The temporal correlation between the spiked upflows and downflows suggests a common source for such events.
For example, the first up-flow peak at $t=172$~min is due to the emergence of the flux tube into the corona, and the down-flow peak at a slightly later time shows the draining following the emergence. The two profound spikes of upflows and downflows will be discussed later.

\begin{figure}
    \centering
    \includegraphics[width=\linewidth]{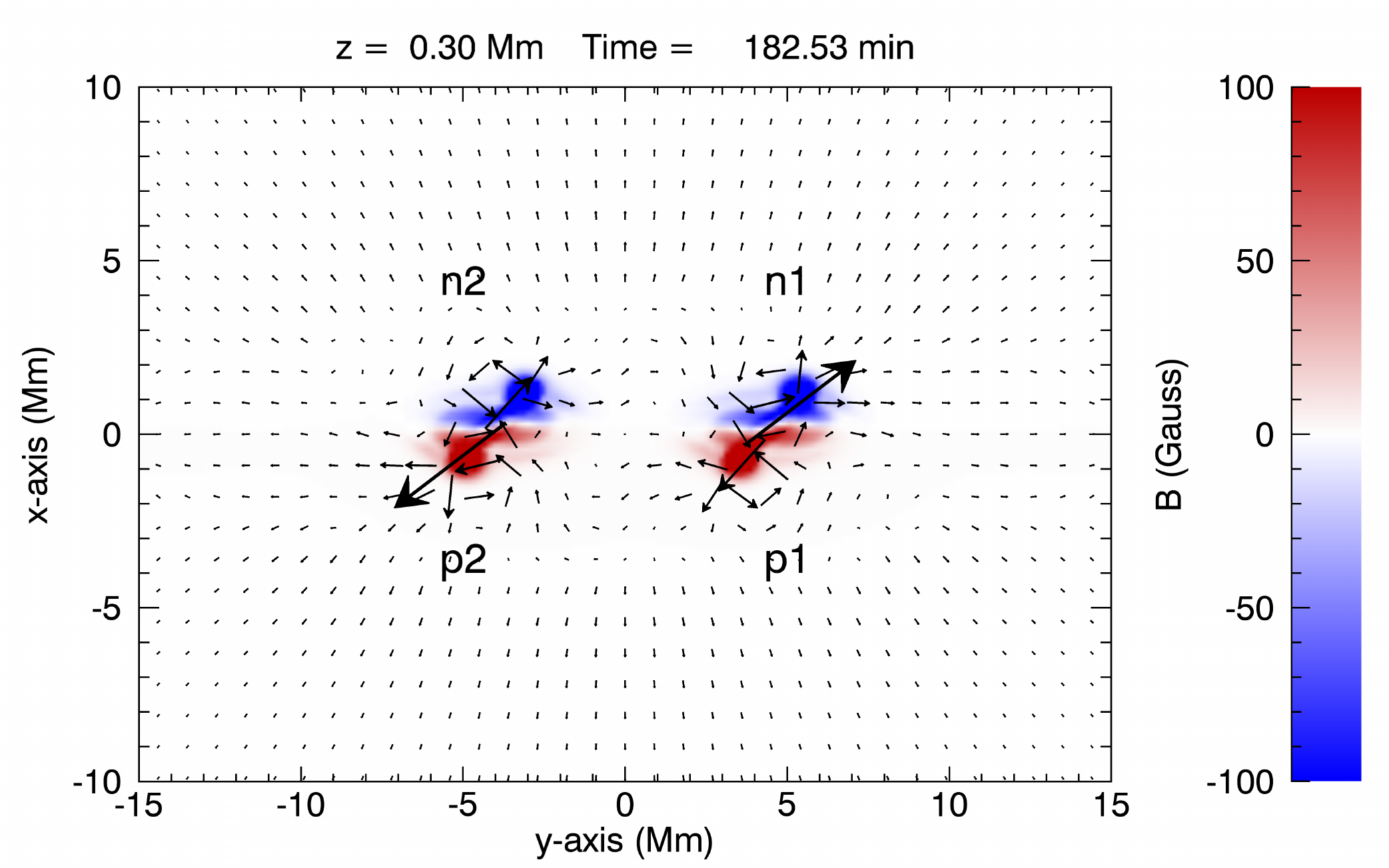}
    \includegraphics[width=\linewidth]{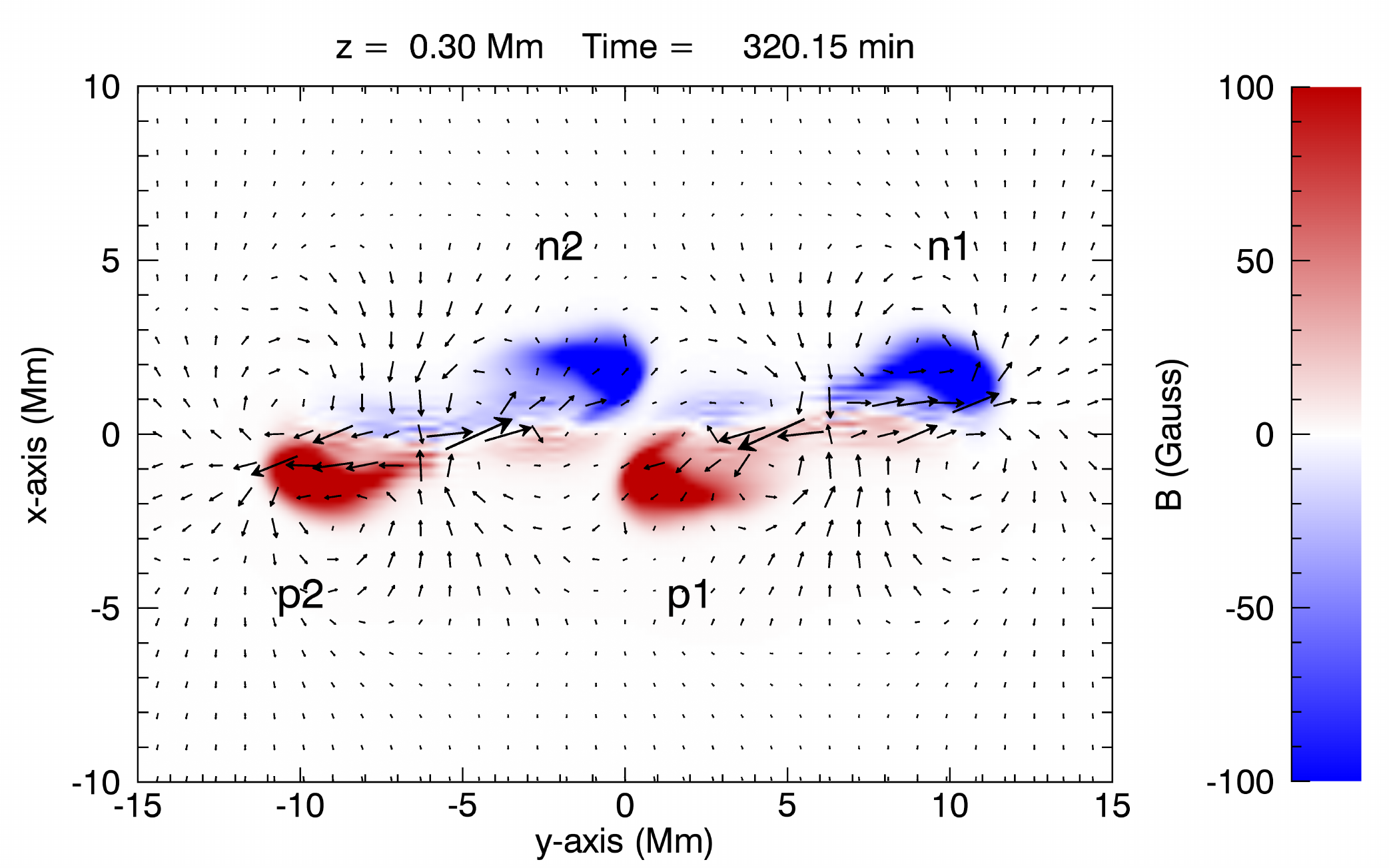}
    \caption{
        The photospheric $B_z$ (red-blue) and the horizontal velocity (arrows).
        The upper panel shows the two bipole shortly after the emergence ($t = 182.5\,\mathrm{min}$). The lower panel shows a later stage of the emergence when the two bipoles have become more sheared ($t = 320.0\,\mathrm{min}$)
    }
    \label{fig:magnetogram}
\end{figure}

The photospheric magnetic field during the evolution of the system shows initially two well-separated bipolar regions (e.g. upper panel, Fig.~\ref{fig:magnetogram}). 
For clarity, we name the two bipoles as bipole 1 (right) and bipole 2 (left). 
Positive $B_z$ (red) is denoted by P whilst the negative (blue) is denoted by N.
The projected velocity field (arrows) indicates that i) the polarities of each bipole diverge away from each other and ii) that converging motions develop at their internal PIL.
The two bipoles move away from the location of their initial emergence, mostly moving along the $y$-axis. Eventually, the (P1, N2) polarities come in contact, forming a new PIL in between them (e.g. lower panel, Fig.~\ref{fig:magnetogram}). 
Hereafter, the pair (P1, N2) is referred to as \textit{inner polarities} and the pair (P2,N1) as \textit{outer polarities}.
Over time, field lines closer to the flux tube axis emerge above the photosphere (the axis of the flux tube always remains below the photosphere). These field lines inject more horizontal field, and therefore more shear, into the photosphere, and as a result contribute to the stressing of the atmospheric field.

\begin{figure}
    \centering
    \includegraphics[width=\linewidth]{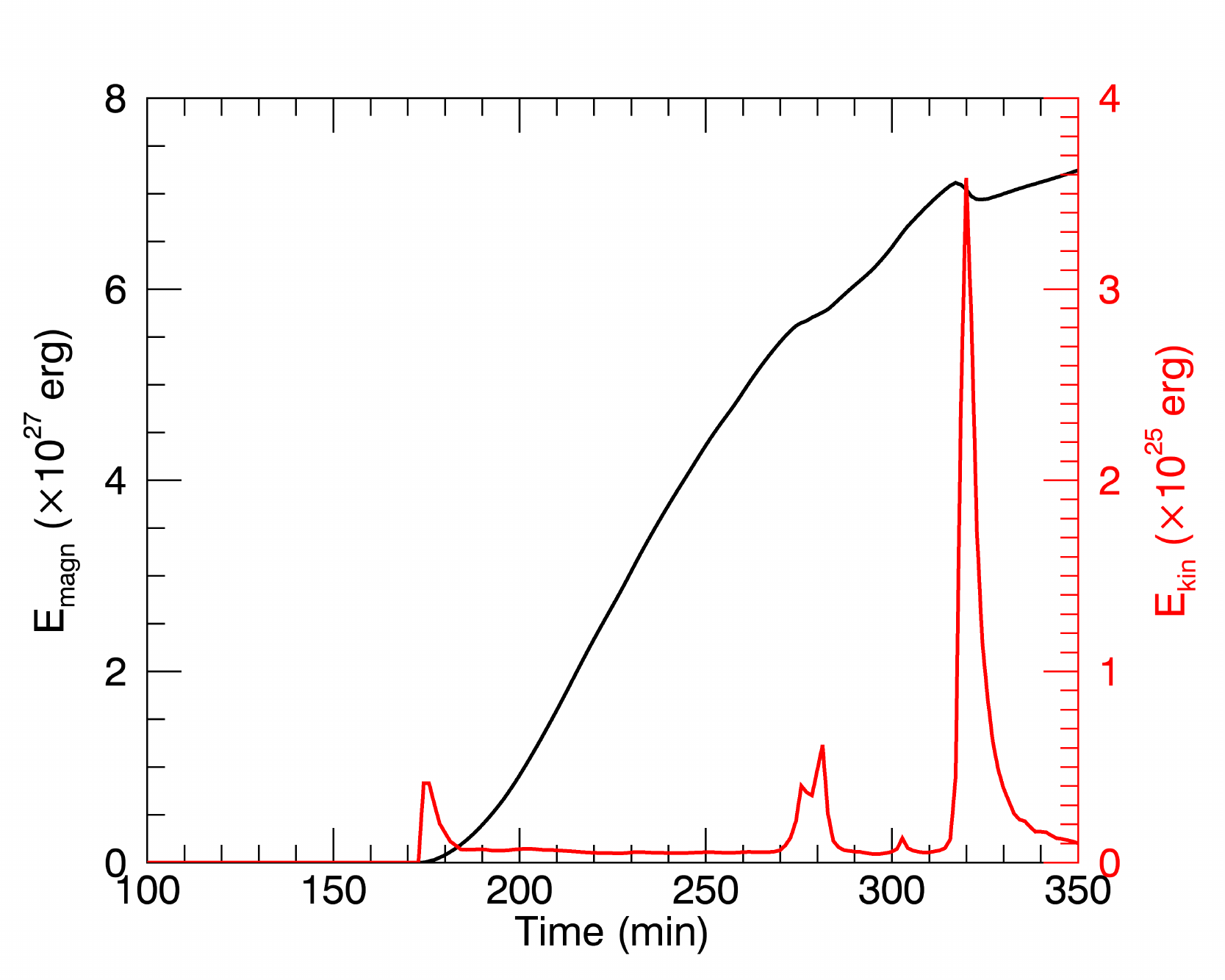}
    \caption{
        Temporal evolution of total magnetic (black) and total kinetic energy (red) inside the corona. 
    }
    \label{fig:energy_time}
\end{figure}

Fig.~\ref{fig:energy_time} shows the total coronal magnetic energy (black) and the total coronal kinetic energy (red).  The magnetic energy increases throughout the simulation as flux continuously emerges into the corona. The magnetic energy starts to saturate towards the end of the simulation, signifying the end of the emergence phase. 
The kinetic energy exhibits two strong peaks coinciding with the profound spikes in $v_{z}$ (Fig.~\ref{fig:vz_time}). These are associated with two flux rope eruptions. During the eruptions, the rate of magnetic energy build-up decreases as free magnetic energy is converted into kinetic energy and heat. The modulation in magnetic energy is of the order of $10^{26}\,\mathrm{erg}$. while the kinetic energy of the eruptions is $0.6-3.6\times10^{25}\,\mathrm{erg}$.

\subsection{First Confined Eruption}
\label{subsec:first_eruption}

\begin{figure*}
    \centering
    \begin{minipage}{0.93\textwidth}
    \centering
    \includegraphics[width=0.9\linewidth]{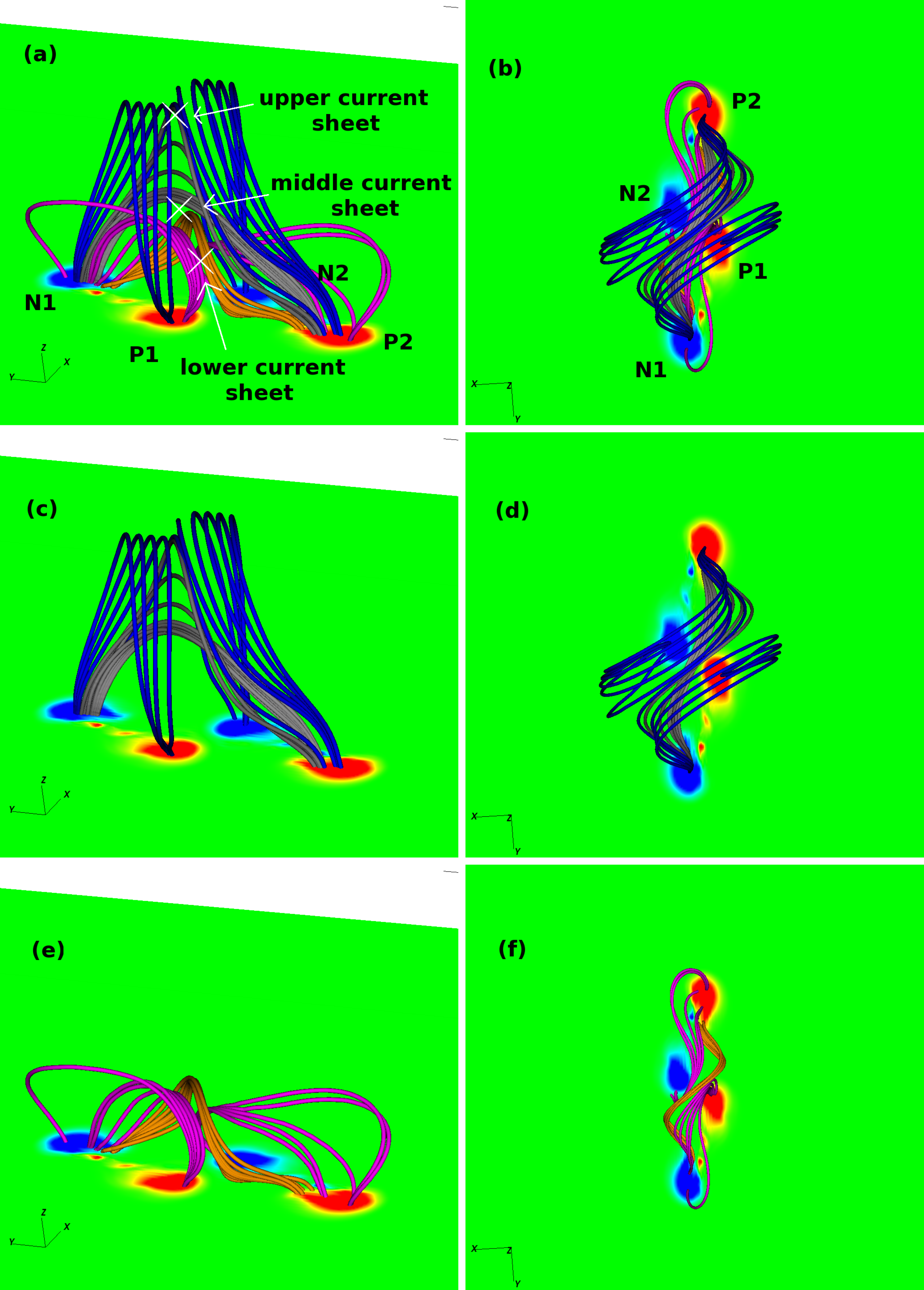}
    \caption{
        The overall magnetic field line topology before the first eruption $t=274$~min ((a) side view, (b) top view). 
        The white crosses indicate the locations where the current sheets are formed. 
        Panels (c) and (d) focuses on the upper part of the field line system. 
        Blue field lines are part of the magnetic field lobes traced around $z=13$~Mm.
        The grey field lines result from the reconnection between blue field lines. 
        Panels (e) and (f) show the lower part of the magnetic field line system. 
        Pink field lines are lower-lying sheared field lines of the magnetic lobes. 
        The orange field lines result from the reconnection between pink field lines. 
        The blue-red patches at the photospheric plane show the negative-positive $B_z$ component of the magnetic field saturated at $\pm300$~G.
    }
    \label{fig:fieldlines1}
    \end{minipage}
\end{figure*}

\begin{figure}
    \centering
    \includegraphics[width=\linewidth]{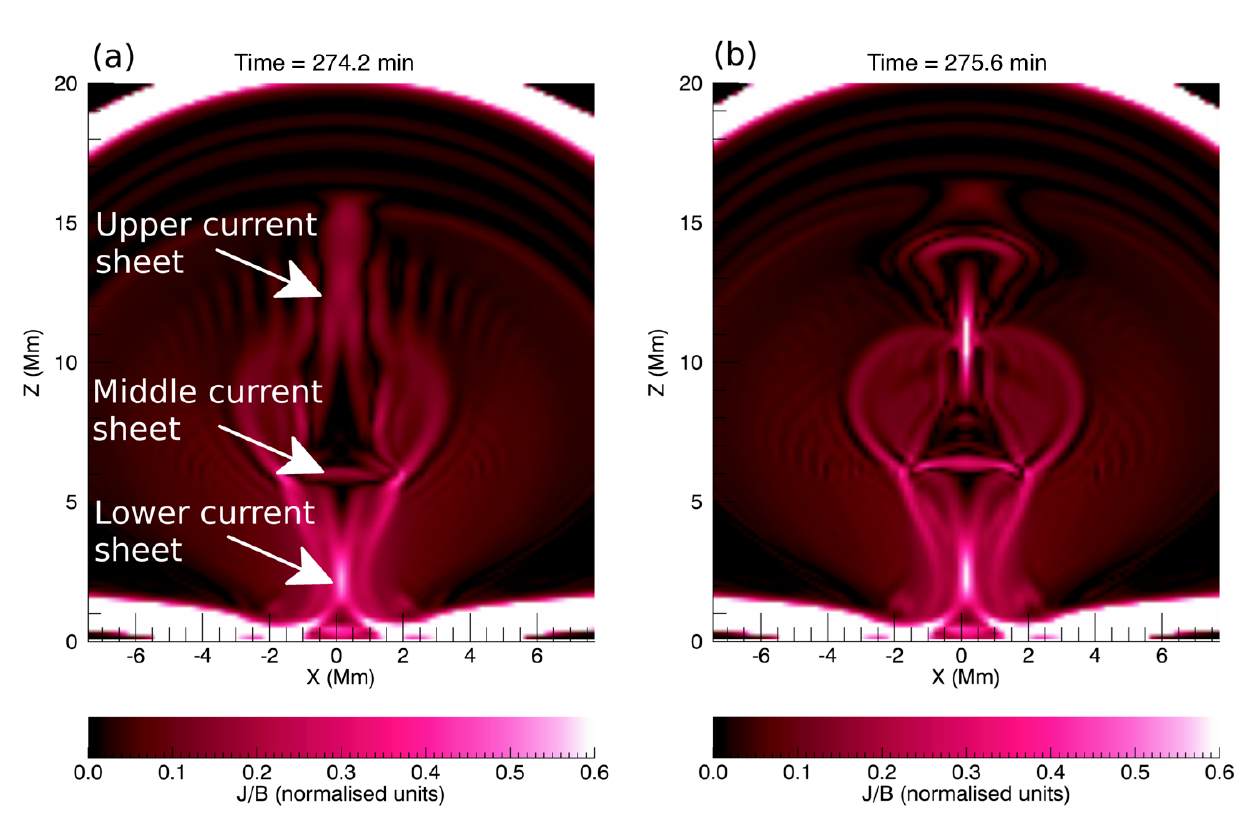}
    \caption{
        $|J/B|$ at the $xz$-midplane of the numerical domain, showing the lower, middle and upper current sheets at (a) $t=274.2$~min and (b) $t=275.6$~min.
    }
    \label{fig:joverb}
\end{figure}

We now focus on the first eruption. 
As described in the previous section, two magnetic lobes are formed above the emerged region (Fig.~\ref{fig:bcross}b). 
Some example lines of these two magnetic lobes, traced from around $z$=13 Mm, are shown in blue in Fig.~\ref{fig:fieldlines1} ($t=274$~min, before the first eruption), connecting (P1, N1) and (P2, N2).
It is clear that the lobes compress against each other due to their lateral expansion (panels a, b).
This 3D expansion results in the formation of a series of current sheets at the interface between the two lobes (e.g. $|J/B|$ at the $xz$-midplane, Fig.~\ref{fig:joverb}a).
A long and extended current sheet is formed higher up in the atmosphere (``upper current sheet''). It extends between $10-16$~Mm and is approximately parallel to the inner polarity inversion line (PIL). Because of the 3D expansion of the lobes, compression is maximum around these heights, while it decreases above and below them.
Reconnection of the lobe field lines along the upper current sheet forms field lines connecting the outer polarities (P2, N1) (grey lines, \ref{fig:fieldlines1}c, d).
These grey lines relax downwards, contain a small amount of shear and twist, and eventually adopt an almost straight configuration. Besides the grey lines, reconnection of the lobe field lines form lines connecting the inner polarities, which will be discussed in the next section.

The field lines of the two magnetic lobes located lower in the atmosphere adopt a J-like shape (pink field lines, \ref{fig:fieldlines1}a, b), due to strong shearing and rotation at their foot-points and due to the emergence of more horizontal field as flux emergence continues \citep[e.g.][]{Syntelis_etal2017}. 
Another current sheet is formed at the interface between the J-like field lines (``lower current sheet'',  Fig.~\ref{fig:joverb}a). Reconnection between the pink lines form the field lines connecting the outer polarities (orange , panels \ref{fig:fieldlines1}e, f) and an arcade connecting the inner polarities (not shown). These orange field lines, being a product of reconnection between sheared lines, are twisted and form a magnetic flux rope. 

\begin{figure}
    \centering
    \includegraphics[width=\linewidth]{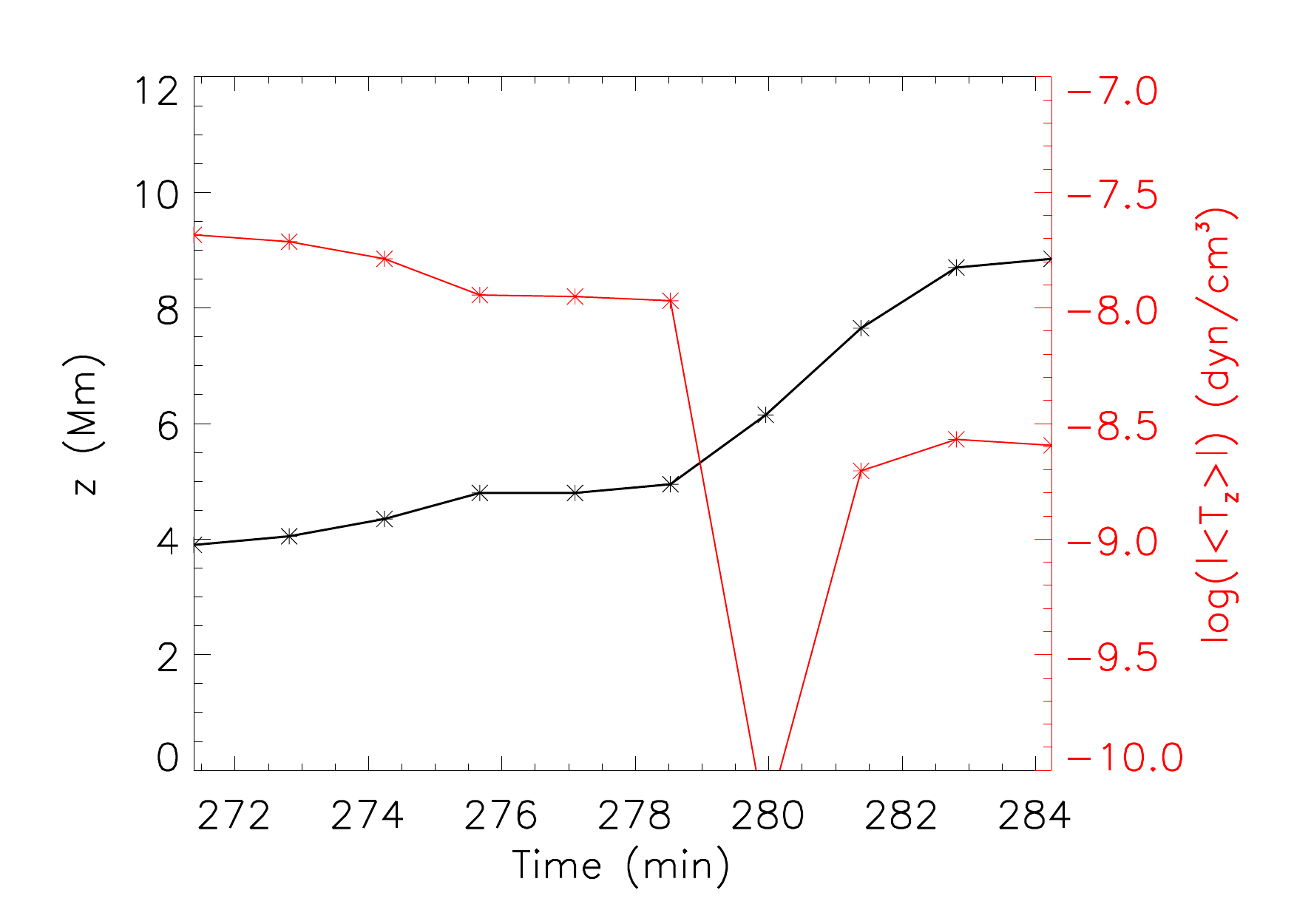}
    \caption{
        The height-time profile for the first flux rope (black line) and the average $\log(|T_z|)$ above the middle current sheet (red line).
    }
    \label{fig:ht1_tension}
\end{figure}

\begin{figure}
    \centering
    \includegraphics[width=\linewidth]{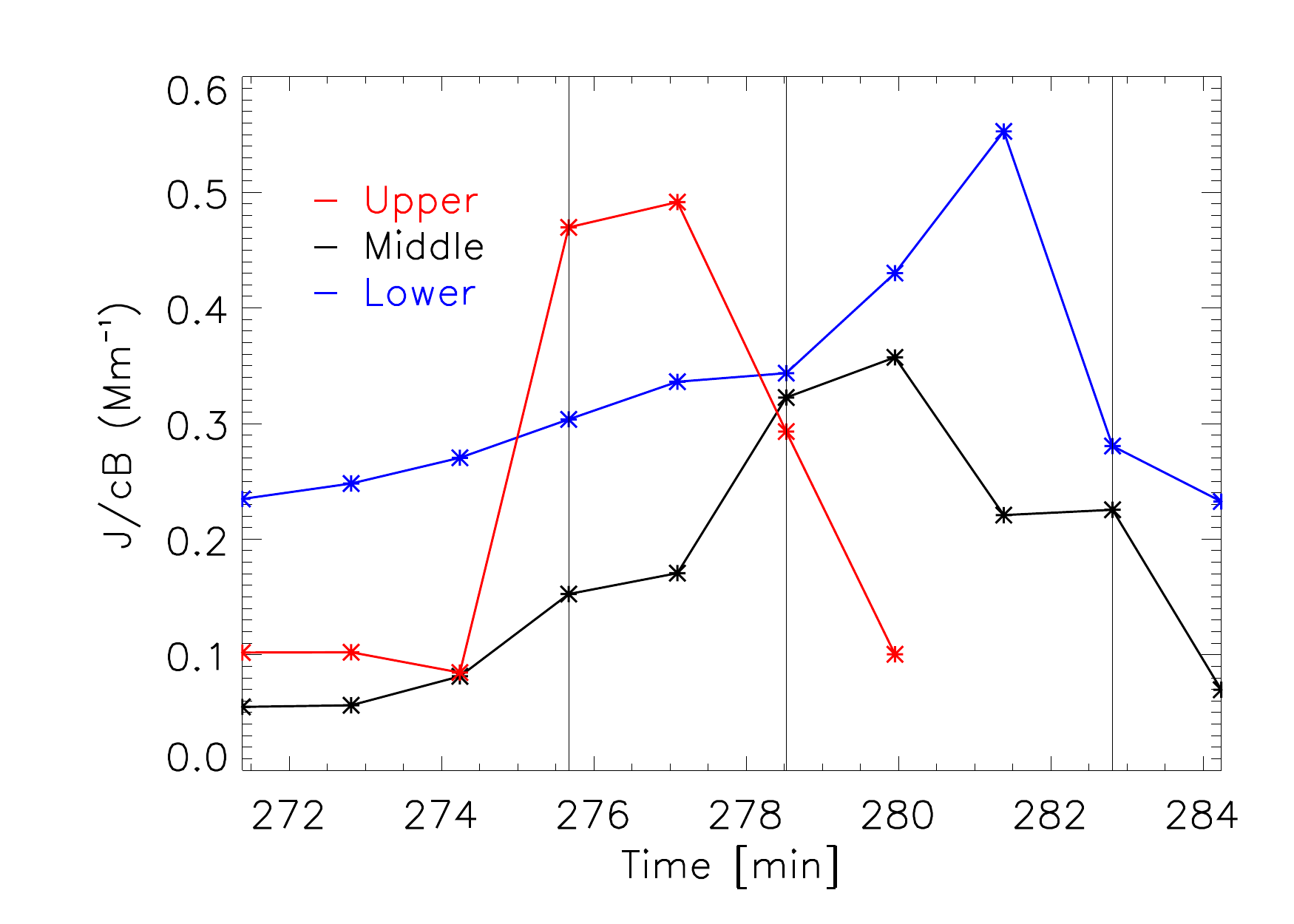}
    \caption{
        Temporal evolution of $J/(cB)$ at the lower current sheet (blue), middle current sheet (black) and  upper current sheet (red). 
        The vertical lines delimit the characteristic phases of the flux rope evolution.
    }
    \label{fig:jb1}
\end{figure}

\begin{figure}
    \centering
    \includegraphics[width=\linewidth]{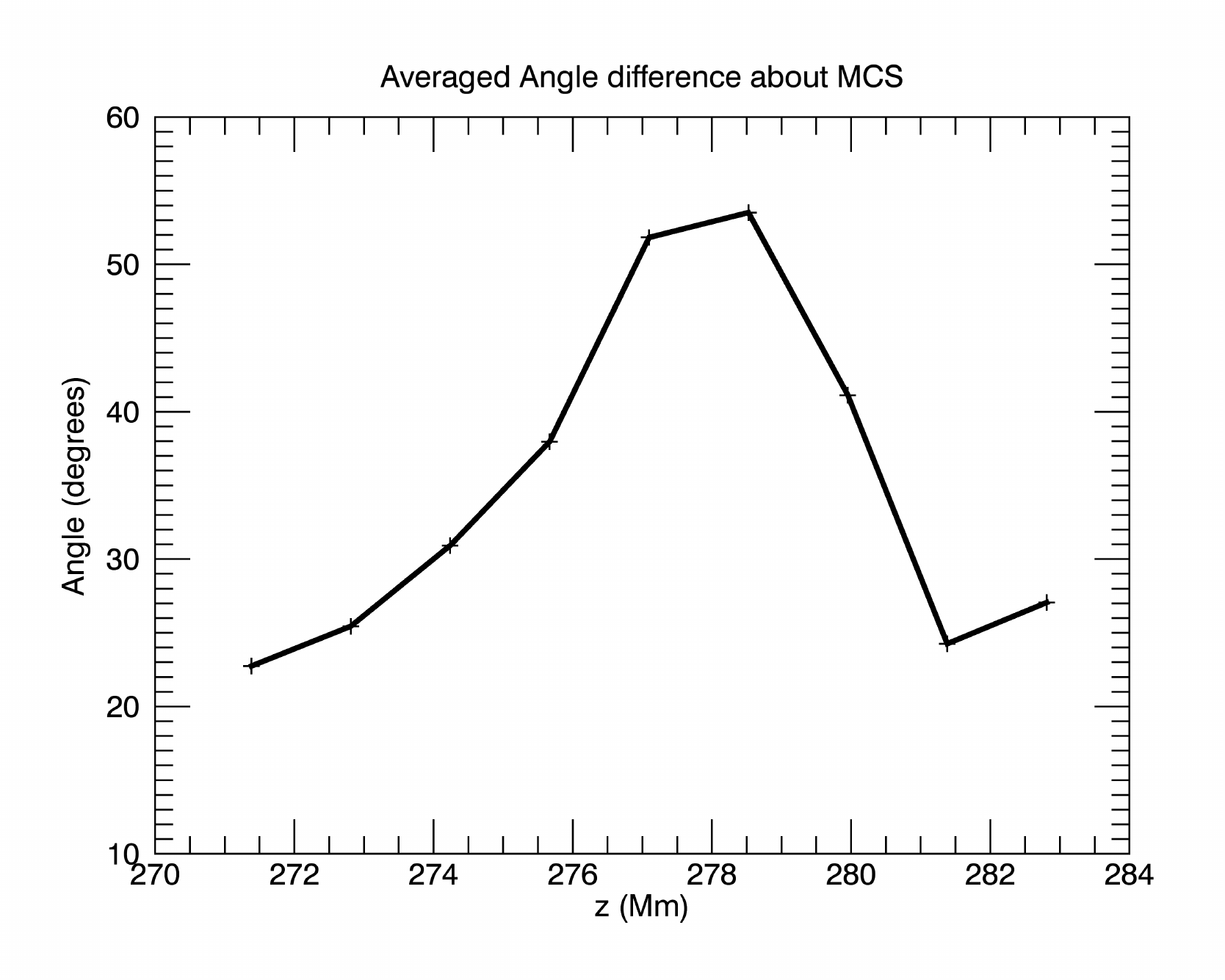}
    \caption{
        The average angle difference between the field lines of the first flux rope (orange lines, Fig.~\ref{fig:fieldlines1}f) and the field lines directly above it (grey lines, Fig.~\ref{fig:fieldlines1}d). The average difference angle is measured between lines of the apex of the flux rope and field lines above the middle current sheet.
    }
    \label{fig:b_angle}
\end{figure}

\begin{figure*}
    \begin{minipage}[b]{0.95\textwidth}
    \centering
    \includegraphics[width=0.9\linewidth]{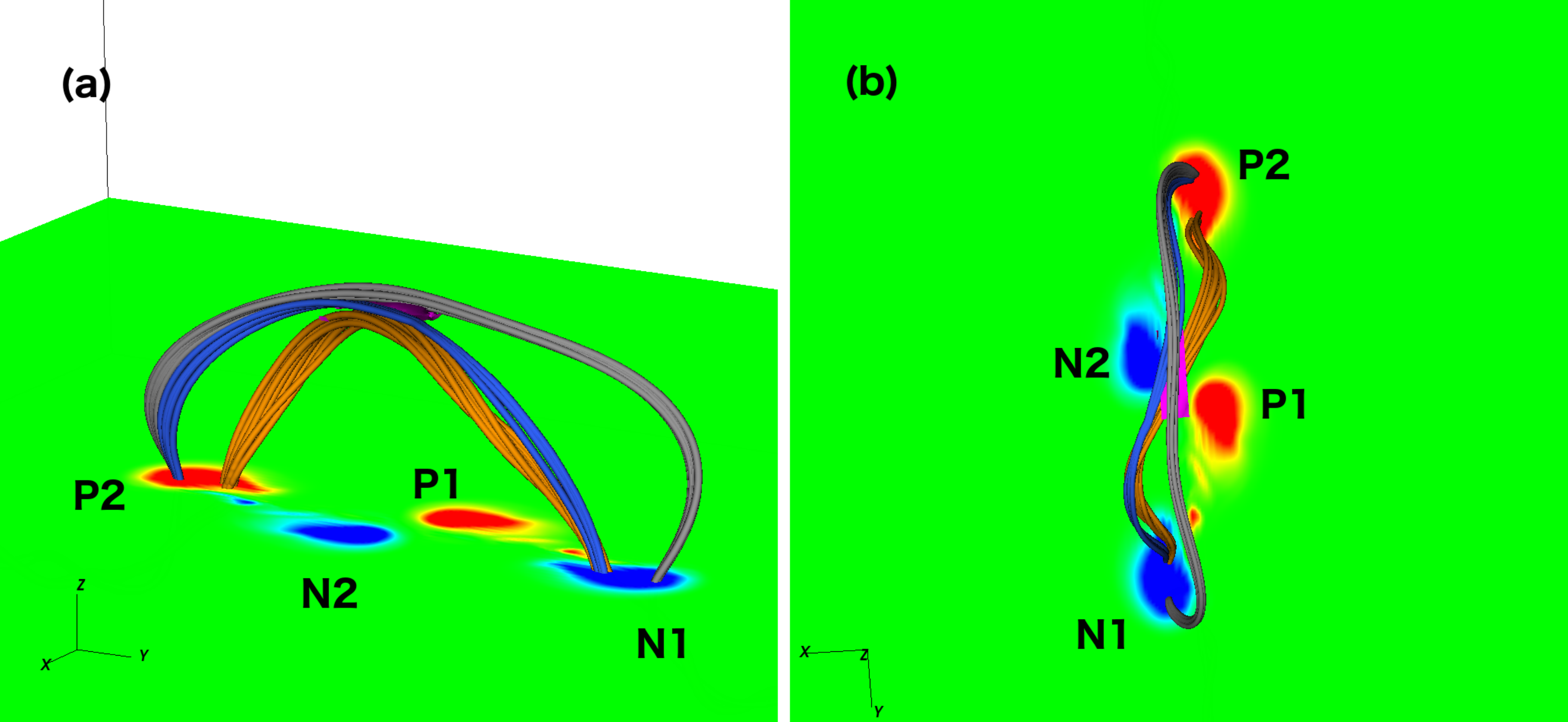}
    \caption{
            Side (a) and top (b) view of the magnetic field line topology at $t=281.4$~min, showing the reconnection of field lines between the flux rope and the overlying field. Purple isosurface shows the middle current sheet.
        }
    \label{fig:reconnection1}
    
    \centering
    \includegraphics[width=0.9\linewidth]{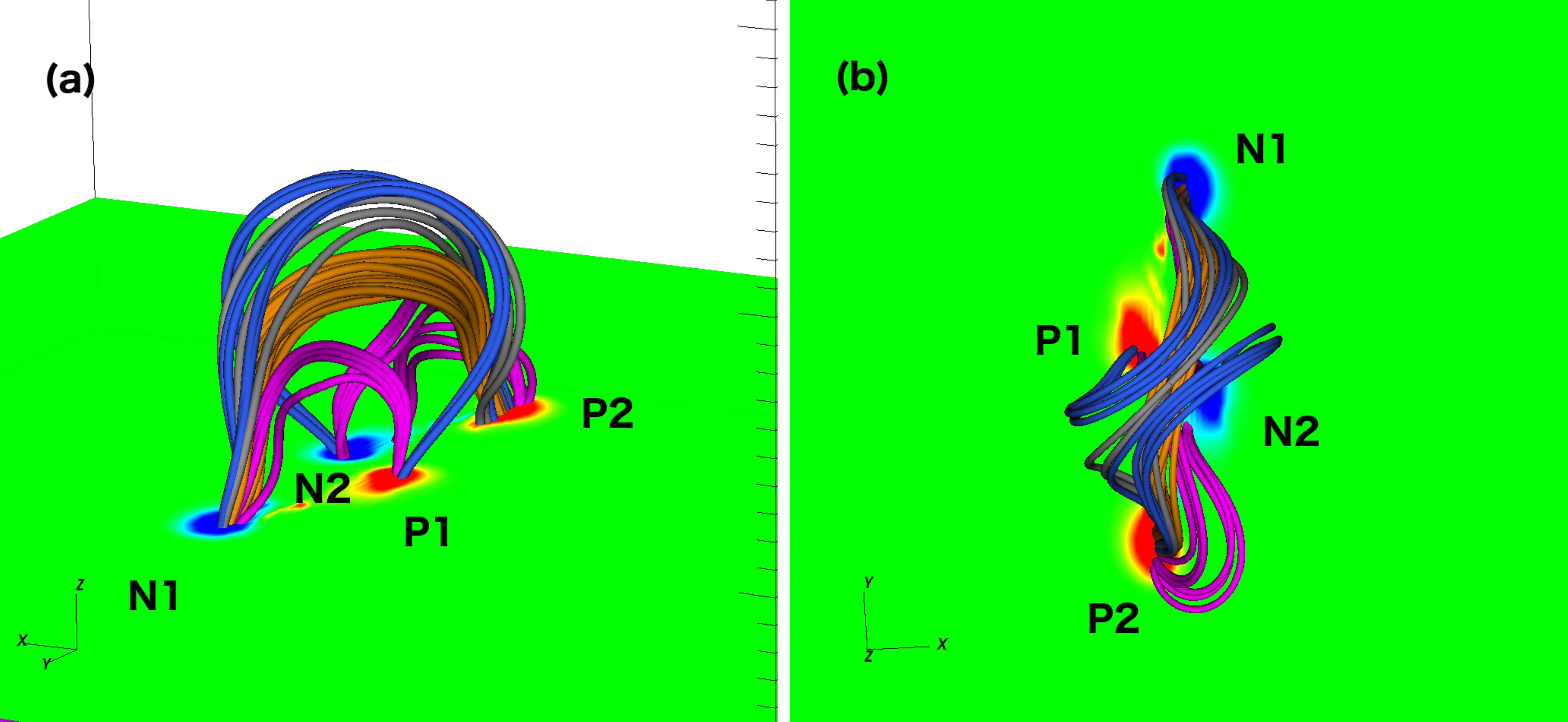}
    \caption{
            Side (a) and top (b) view of the magnetic field line topology at $t=283$~min, after the first eruption.
        }
    \label{fig:eruption1}
    \end{minipage}
\end{figure*}

\begin{figure*}[ht!]
    \centering
    \begin{minipage}[b]{0.95\textwidth}
    \centering
    \includegraphics[width=0.8\linewidth]{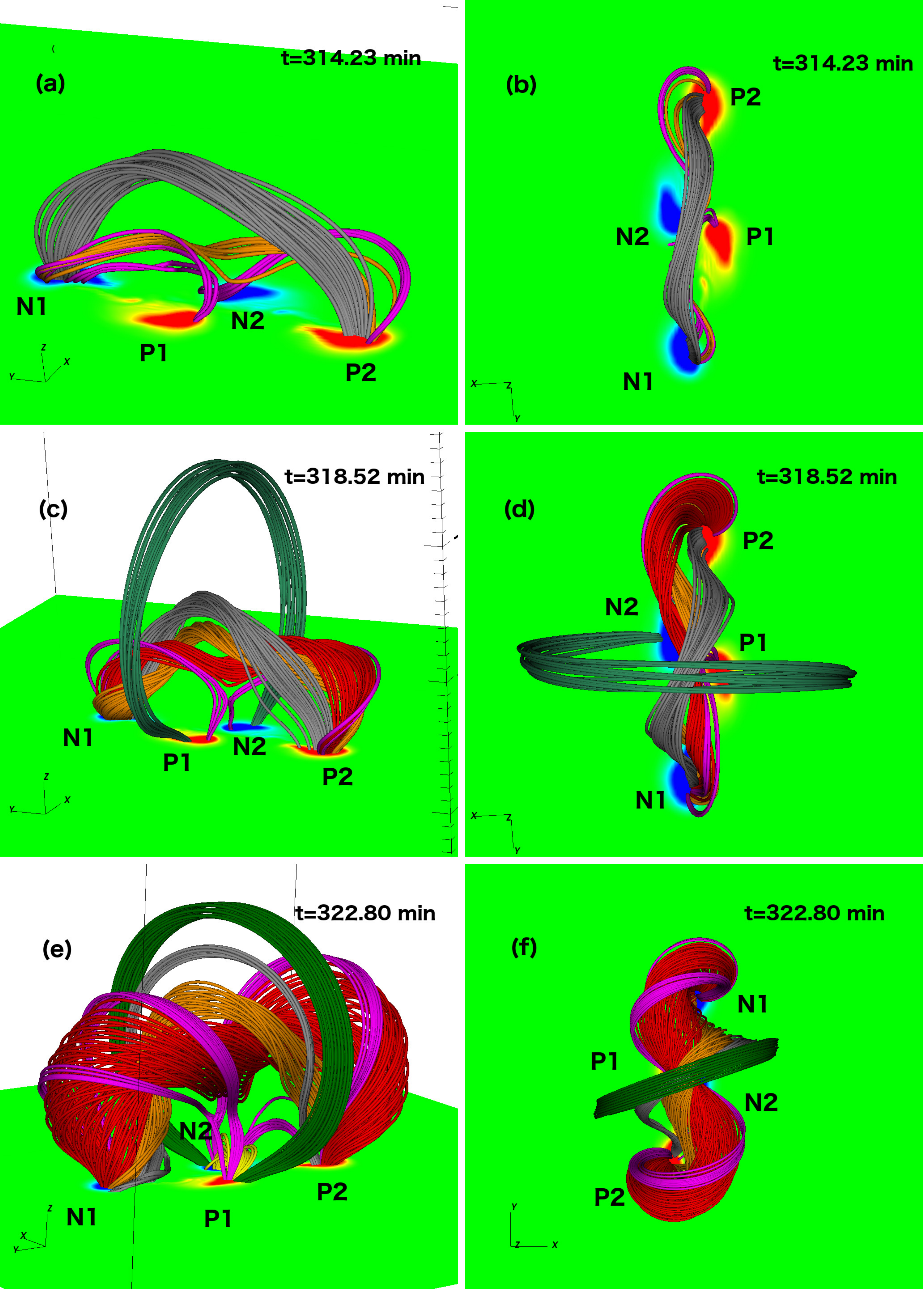}
    \caption{
            Side (a) and top (b) view of the magnetic field line topology of the second flux rope before its eruption ($t=314$~min). 
            The field lines are similar to the previous eruption and are colored as in Fig.~\ref{fig:fieldlines1}. 
            Panels (c) and (d) show the flux rope during the eruption at $t=318.5$~min. 
            Green field lines result from the reconnection of field lines similar to the blue ones of Fig~\ref{fig:eruption1}a, and act as a strapping field to the eruption.
            Red field lines are formed by the reconnection of pink lines during the eruption of the flux rope core (orange).
            Panels (e) and (f) show the magnetic field line topology at $t=322.8$~min, when the eruption has stopped and the flux rope has been confined. 
            The yellow field lines connecting the inner polarities (P1,N2) are the post-reconnection arcade loops.
    }
    \label{fig:fieldlines2}
    \end{minipage}
\end{figure*}

Note that both the orange and the grey lines are formed by the reconnection of the field lines of the two magnetic lobes. However, because the reconnection occurs at different heights, the lines that reconnect (blue/pink) have different amount of shear and different orientation. 
Therefore, the resulting grey/orange lines also have different amount of shear and twist and different orientation (e.g. relative angle between grey/orange lines in Fig.~\ref{fig:fieldlines1} d, f).
The collapsing grey field lines press down on top of the flux rope generating a ``middle current sheet'' (Fig.~\ref{fig:joverb}a, Fig.~\ref{fig:fieldlines1}a). 

The flux rope core is associated with cooler temperatures, higher densities and a dominate axial magnetic field component. 
We therefore identify the flux rope's axis by examining temperature, density and axial magnetic field, across 2D vertical planes, perpendicular to the flux rope \citep[similar to][]{Syntelis_etal2017}. 
The height-time profile of the flux rope is shown in Fig.~\ref{fig:ht1_tension} (black). 
The flux rope's height-time profile exhibits three phases: 
i) a slow rise phase between $271.5 - 275.6\,\mathrm{min}$, 
ii) a temporary inhibition of the slow rise between $275.6 - 278.5\,\mathrm{min}$, and 
iii) an eruptive phase leading to a confinement between $278.5 - 284.2\,\mathrm{min}$

These rise phases are closely related to reconnection occurring at the current sheets of the system. Thus, we follow the temporal evolution of the maximum $J/(cB)$ at the lower (blue), middle (black) and upper (red) current sheets (Fig.~\ref{fig:jb1}). During the first phase (until first vertical line), the flux rope rises due to magnetic pressure gradient force. 
As the flux rope pushes upwards against the overlying grey field lines, the middle current sheet is enhanced.

During the second phase, between $t = 275.6 - 278.5\,\mathrm{min}$, the upper current sheet exhibits a sudden bust of reconnection (red line, Fig.~\ref{fig:jb1}, and Fig.~\ref{fig:joverb}b). The long and thin upper current sheet becomes fragmented, forming two regions of increased current density and triggering fast reconnection flows.  This causes the first of the two peaks associated with the $v_z$ and kinetic energy increase between $t = 275.6 - 278.5\,\mathrm{min}$ in Fig.~\ref{fig:vz_time} and Fig.~\ref{fig:energy_time} respectively. This fragmentation is suggestive of a tearing instability. 
This sudden reconnection results in the formation of more grey field lines overlying the flux rope, and also temporarily increases the axial flux above the rope between $t = 275.6 - 278.5\,\mathrm{min}$. These newly formed grey lines retract downwards and inhibit the rise of the flux rope between $t = 275.6 - 278.5\,\mathrm{min}$. The average downwards tension above the middle current sheet (red, Fig.~\ref{fig:ht1_tension}) before  $t = 275.6$~min  is decreasing gradually due to the expansion of the magnetized volume. Between $t = 275.6 - 278.5\,\mathrm{min}$, however, this gradual decrease of the tension stops due to the retracting grey lines, indicating how these lines inhibit the rise of the flux rope. 

As the flux rope push upwards and the grey field lines push downwards, the middle current sheet becomes enhanced.
In addition to that, between $271.5 - 278.5\,\mathrm{min}$ as the flux rope i) becomes more sheared and ii) moves upwards, the average relative angle between the apex of the flux rope and the grey field lines above the middle current sheet increases (Fig.~\ref{fig:b_angle}), with the maximum relative angle taking values around 90$^{\circ}$. 
Eventually, reconnection between the field lines of the apex of the flux rope and the grey field lines above the middle current sheet becomes more efficient, marking the beginning of the eruption of the first flux rope at $278.5\,\mathrm{min}$ (second of the two peaks associated with the $v_z$ and kinetic energy increase between $t = 275.6 - 278.5\,\mathrm{min}$ in Fig.~\ref{fig:vz_time} and Fig.~\ref{fig:energy_time} respectively).
Reconnection intensifies at the middle current sheet, reducing the magnetic tension of the field above it ($t = 280\,\mathrm{min}$, red line, Fig.~\ref{fig:ht1_tension}).
This process is visualised in Fig.~\ref{fig:reconnection1}a,b where the flux rope (orange) and the overlying field (grey) reconnect through the middle current sheet (purple isosurface) to form the blue lines. These newly formed blue lines have less downwards tension and are shifted away from the flux rope apex, aiding the eruption. 
As the flux rope erupts, the lower current is also enhanced.

Eventually, most of the flux rope's field reconnects with the field above it and the eruption stops. The resulting field after the eruption contains a structure (orange, Fig.~\ref{fig:eruption1}a,b), consisted of some remaining flux rope lines, some grey lines and some of the lines formed from the reconnection between the two.

\begin{figure}
    \centering
    \includegraphics[width=\columnwidth]{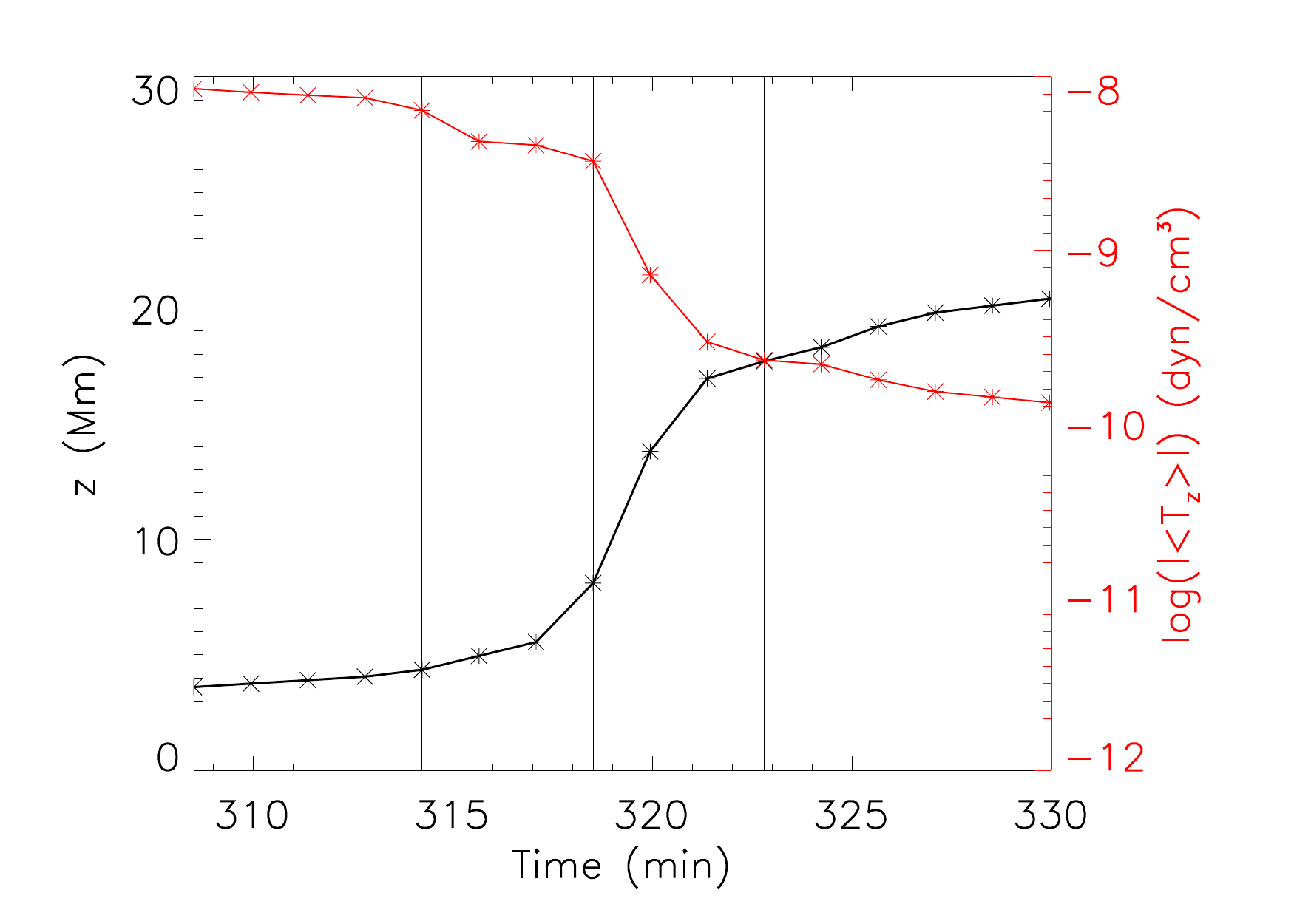}
    \caption{
       The black line shows the height-time profile of the second flux rope. Vertical lines indicate the beginning of the gradual acceleration phase, the beginning of the eruptive phase and the confinement phase. The red line shows the temporal evolution  of  the  average  absolute  magnetic  tension  in  a cross-sectional cut area above the flux rope.
    }
    \label{fig:ht2_tension}
\end{figure}

\begin{figure}
    \centering
    \includegraphics[width=\linewidth]{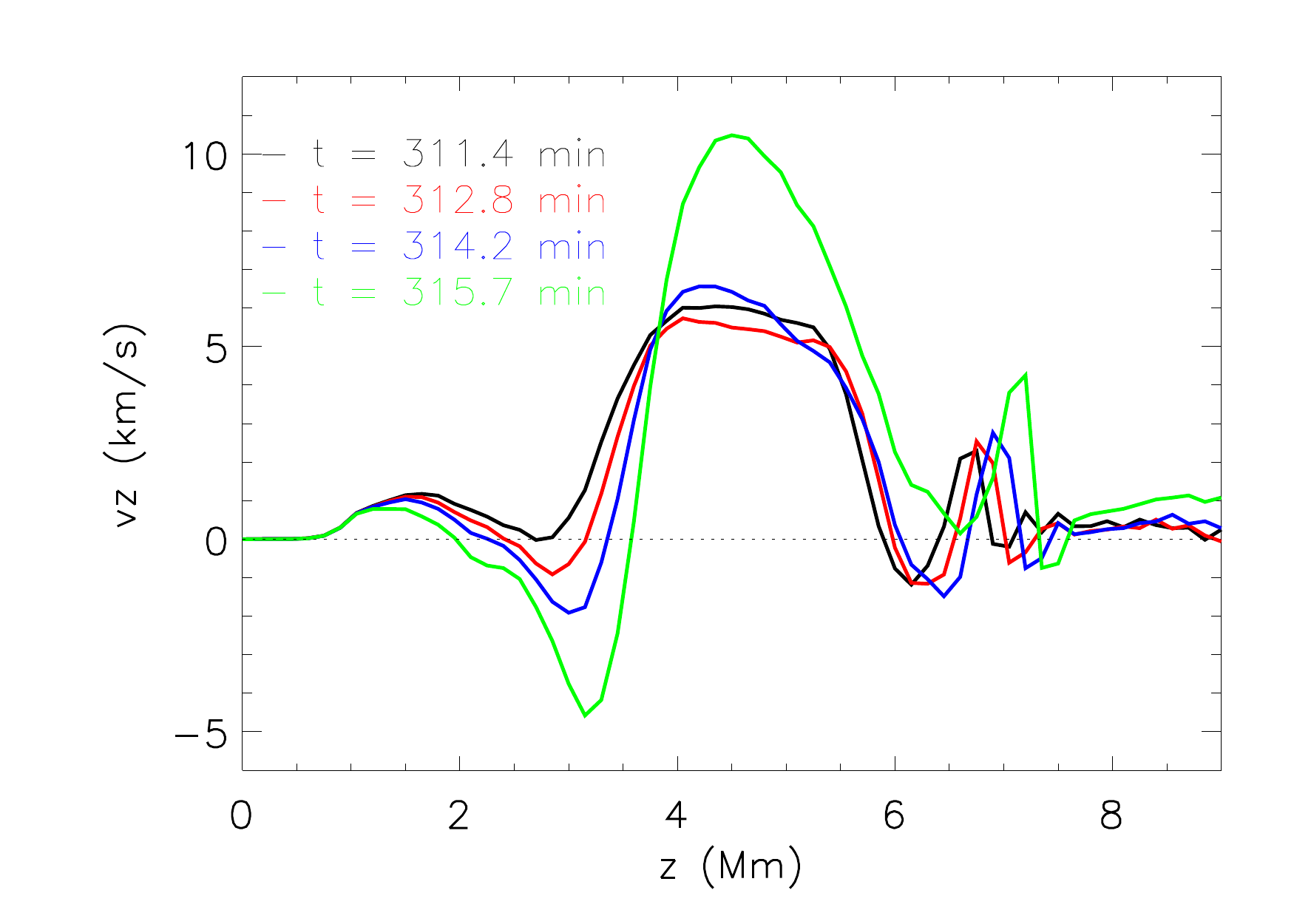}
    \includegraphics[width=\linewidth]{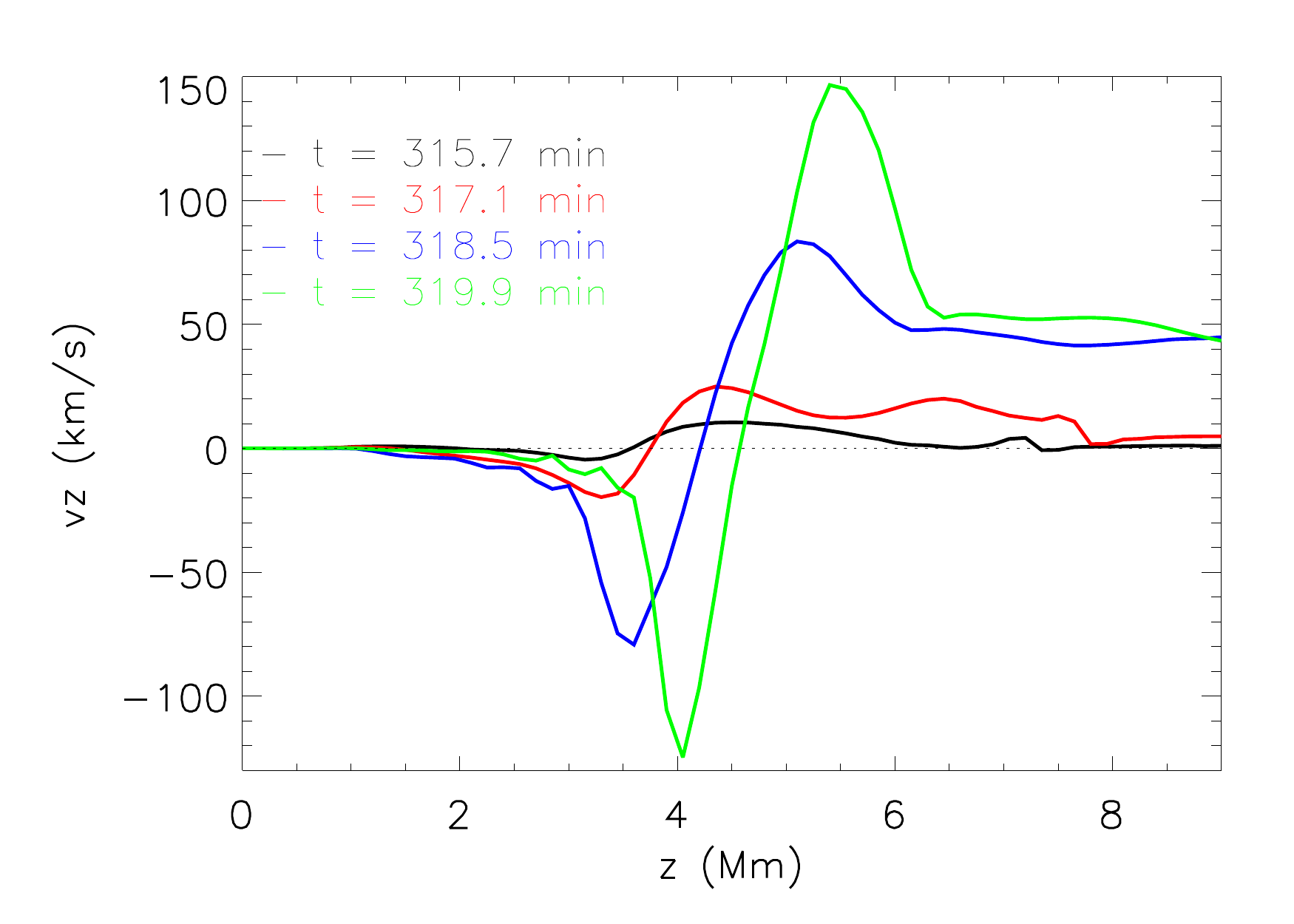}
    \caption{
        $v_{z}$ along $(0,0,z)$ at different times denoted by colour. The upper panel shows earlier times and the lower panel is the continuation at later times. 
        We see strong bidirectional jets develop as reconnection drives an outflow from the lower current sheet.
    }
    \label{fig:vz}
\end{figure}

\begin{figure}
    \centering
    \includegraphics[width=\columnwidth]{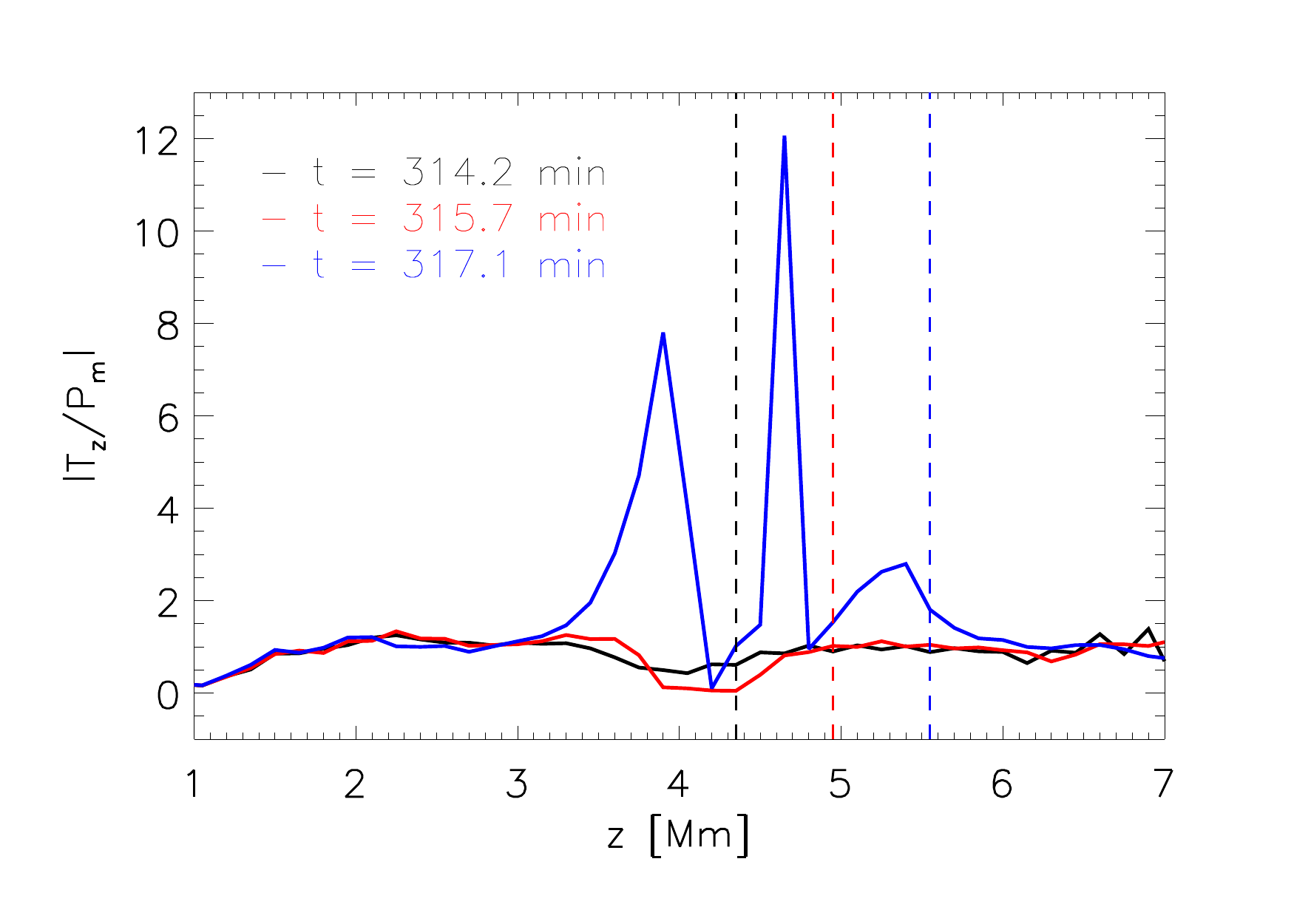}
    \caption{
        The ratio of the absolute value of the vertical component of the magnetic tension force over the magnetic pressure force along $(0,0,z)$ at different times. The vertical dashed lines indicate the location of the flux rope center at these times.
    }
    \label{fig:forces2}
\end{figure}

\begin{figure}
    \centering
    \includegraphics[width=\columnwidth]{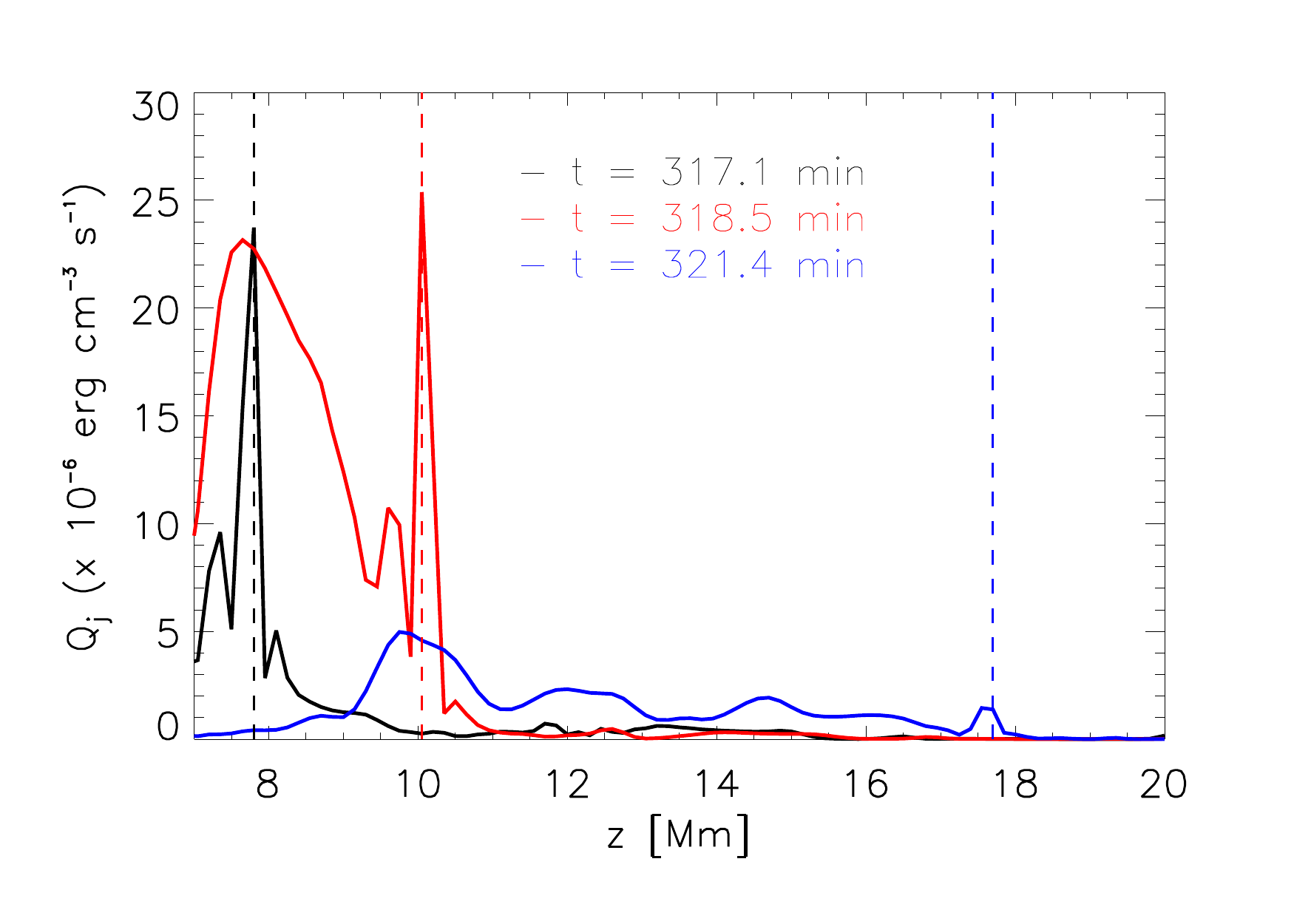}
    \caption{
        Joule heating ($Q_{j}$) along $(0,0,z)$ at different times. The vertical dashed lines indicate the location of the middle current sheet above the flux rope at these times.
    }
\label{fig:qj2}
\end{figure}

\subsection{Second Confined Eruption}
\label{subsec:second_eruption}

We now focus on the second eruption. The pre-eruptive configuration is visualized in Fig.~\ref{fig:fieldlines2}a,b. 
We again identify low-lying J-shaped field lines (pink), which are now more sheared. These lines reconnect, forming a second twisted flux rope core (orange).
Above the flux rope, field lines connect the outer polarities in a similar manner to the first eruption (grey lines). These grey lines consist of the remnants of the first eruption and the continued retraction of reconnected lines at the the upper current sheet.

We trace the height time profile of the second flux rope in Fig.~\ref{fig:ht2_tension} (black) and over-plot the average tension above the flux rope (red). 
The initial slow rise of the flux rope ($t=308.5-314.2$~min, before first vertical line)  progresses towards a non-linear phase ($t=314.2-317.1$~min, between first and second vertical lines). 
This is then followed by a rapid eruption, triggered at $t = 318.5\,\mathrm{min}$ (after second vertical line). Eventually, the flux rope eruption becomes confined (after third vertical line).

Similar to the first eruption, the reconnection between the flux rope apex field lines and the middle current sheet is enhanced over time, reducing the tension of the overlying field (red line, first vertical line Fig.~\ref{fig:ht2_tension}). This release of tension makes the flux rope move upwards, enhancing the reconnection  at both the lower and the middle current sheet (the bidirectional outflows (upper panel, Fig.~\ref{fig:vz}) from both the lower current sheet (around $z=3$~Mm) and the middle current sheet (around $z=7$~Mm) start to increase from $t = 314.2\,\mathrm{min}$).

As the flux rope accelerates upwards ($t=315.7-319.9$~min), more magnetic field flows into the lower current sheet and reconnects. 
Because this field (e.g. pink lines, Fig.~\ref{fig:fieldlines2}a) is more sheared than in the first eruption, this reconnection now results to new lines that have higher upwards tension. As a result, from $t = 314.2-317.1\,\mathrm{min}$ the ratio of the upwards magnetic tension over the upwards magnetic pressure along height (black/red/blue solid lines, Fig.~\ref{fig:forces2}) shows an increase by a factor of 12 below the flux rope center (indicated by the black/red/blue vertical dashed lines). This upwards tension release from the lower current sheet drives the flux rope upwards and enhances reconnection at the middle current sheet. 
The reconnection rate at the middle current sheet, between field lines of the flux rope apex and field lines above the current sheet, increases over time in a runaway manner.

Eventually, reconnection at the middle current sheet becomes very enhanced ($t=317.1$~min, joule heating peak at the location of the black vertical dashed line, Fig.~\ref{fig:qj2}). This results to a rapid release of the tension of the overlying the flux rope (red line, second vertical line, Fig.~\ref{fig:ht2_tension}), triggering the eruption. 
The corresponding outflows from the lower current sheet increase from $10$ to $150$ km/s (lower panel, Fig.~\ref{fig:vz}), forming a strong reconnection jet that is ejected from the lower (now flare) current sheet. 
The field lines consisting the fast reconnection reconnection outflow have a U-loop shape, further enhancing the upwards tension release below the flux rope (red lines, Fig.~\ref{fig:fieldlines2}), and thus further pushing the flux rope upwards. 
These twisted red field lines wind around the orange flux rope core, enhancing the magnetic flux of the erupting flux rope.

However, similar to the first eruption, the sudden rise of the flux rope does not enhance the middle current sheet sufficiently to remove all the overlying tension. 
Indeed, the Joule dissipation at the middle current sheet does eventually drop off (blue vertical dashed line, Fig.~\ref{fig:qj2}) and consequently, the tension of the overlying field reduces at a slower rate (red line, Fig.~\ref{fig:ht2_tension}). 
In addition, the flux rope does not stretch the overlying field to the extent where tether-cutting reconnection is triggered. 
Eventually, the eruption is halted by the tension of the overlying field (after third vertical line, Fig.~\ref{fig:ht2_tension}) and the flux rope eruption becomes confined by the magnetic envelope (green and grey lines, Fig.~\ref{fig:fieldlines2}e,f)

Fig.~\ref{fig:fieldlines2}e,f shows the field lines of the system after the eruption becomes confined. The flux rope (orange and red) has increased in size and flux and is now more twisted. 
Below the flux rope, a post-reconnection flare arcade connecting the inner polarities has been formed (yellow).
Some of the J-like field lines (pink) have been stretched during the eruption, without reconnecting, and provide tension holding down the flanks of the flux rope.  
Above this, the green and grey overlying field lines act as a strapping field. 
It is worth mentioning here that the green and grey field lines are the two products of the reconnection higher up in the atmosphere between the two original magnetic lobes (blue, Fig.~\ref{fig:fieldlines1}). 
Both of these field line systems should develop naturally in any quardapolar region and can suppress an eruptive flux rope. 
Notice that the green lines form a third magnetic ``lobe'', along with the two ones associated with the two emerged bipoles (blue lobes of Fig.~\ref{fig:fieldlines1}). 
Our analysis show that this structure plays a critical role in the dynamics associated with quadrupolar regions.

\begin{figure*}[ht!]
    \centering
    \begin{minipage}[b]{\textwidth}
    \centering
    \includegraphics[width=0.85\linewidth]{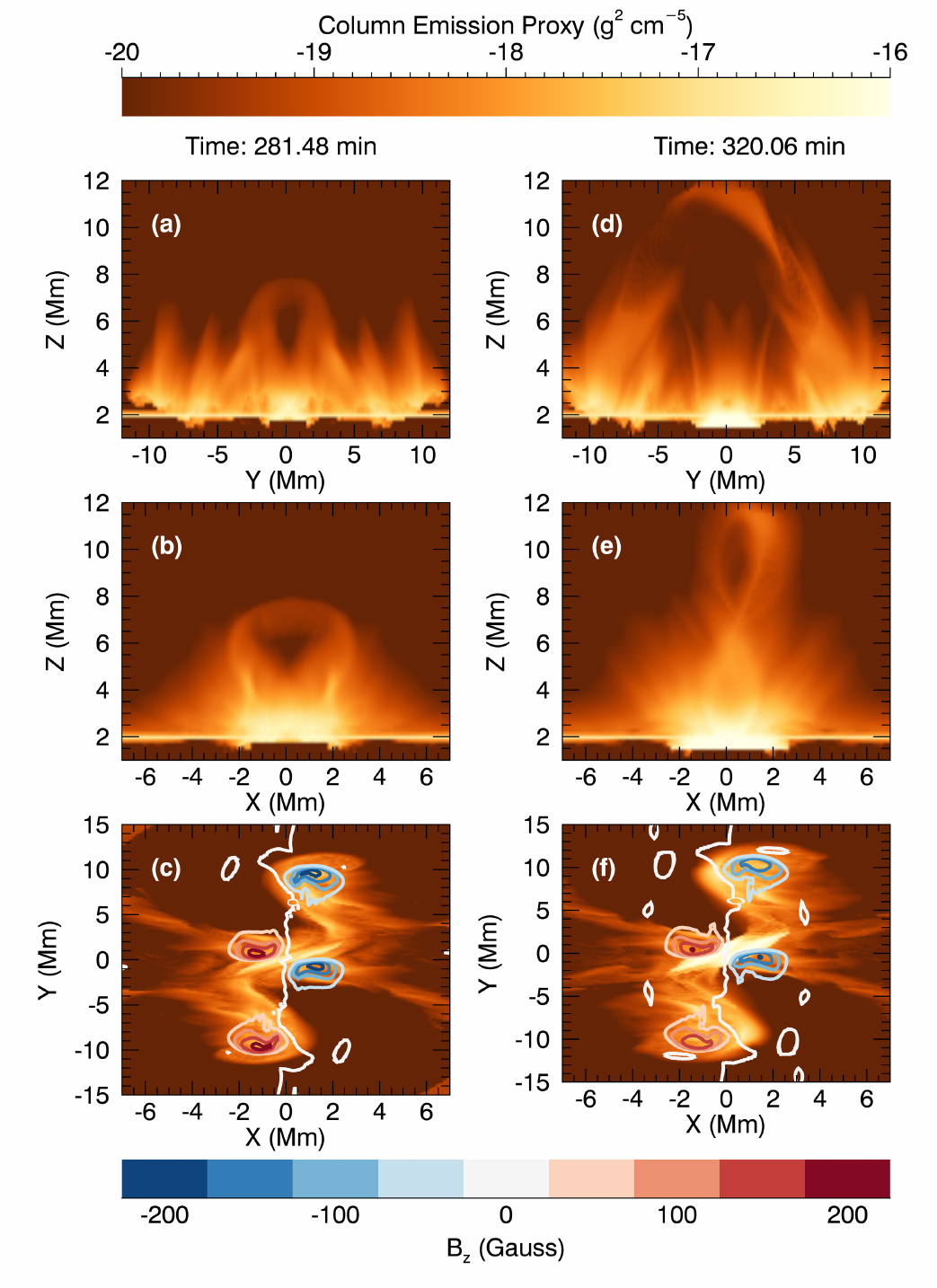}
    \caption{
        Emission proxy images for the cooler temperature range ($30,000 - 300,000\,\mathrm{K}$) for the first eruption at $t = 285.8\mathrm{min}$ (left column) and the second eruption at $t = 320.1\mathrm{min}$ (right column). 
        The upper, middle and lower panels corresponds to integrating over the X, Y and Z lines of sight respectively. 
        The $B_z$ magnetic field contours at the photosphere are overlaid in the lower panel. 
    }
    \label{fig:emiss_cool}
    \end{minipage}
\end{figure*}

\begin{figure*}[ht!]
    \centering
    \begin{minipage}[b]{\textwidth}
    \centering
    \includegraphics[width=0.85\linewidth]{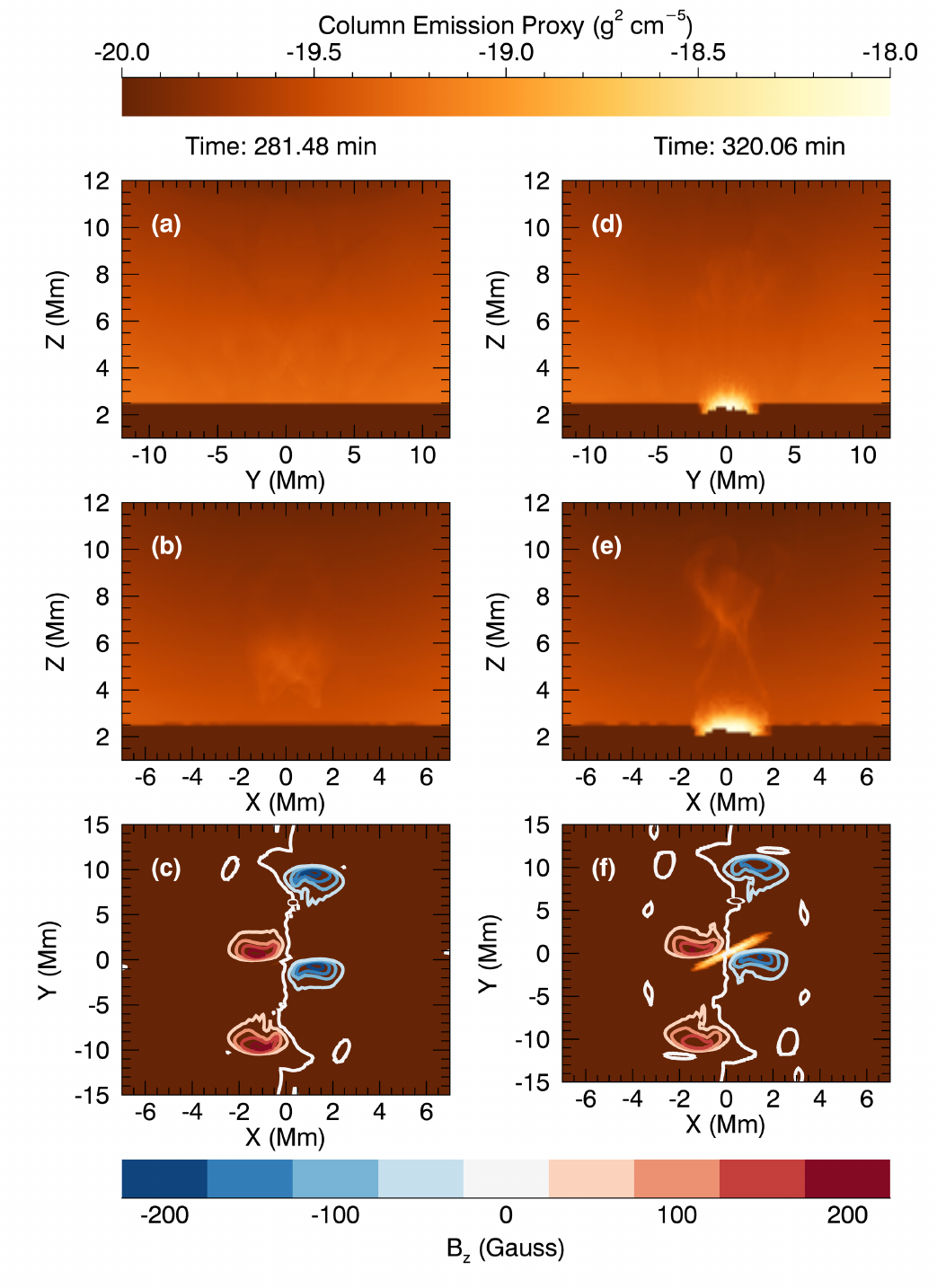}
    \caption{
        Same as Fig.~\ref{fig:emiss_cool} but for the hotter temperature range ($0.6-2$~MK).
    }
    \label{fig:emiss_hot}
    \end{minipage}
\end{figure*}

\subsection{Pseudo-emission and Flaring}
\label{subsec:flare}

We generate an emission proxy ($EP$) to further examine the two eruptions. To do so, we integrate the density of plasma, within a specified temperature interval, over a line of sight:
\begin{equation}
    EP = \int g(T) \rho^{2} ds,
\end{equation}
where $g(T)$ is $1$ inside a selected temperature interval and $0$ outside it. We adopt a temperature interval of $30,000 - 300,000\,\mathrm{K}$ which is typical for cooler plasma in the chromospheric/transition region range, and a temperature interval of $0.6 - 2 \mathrm{MK}$ which is typical for hotter coronal plasma.

The emission proxies are generated during the first (left column) and second (right column) eruption, for the cooler (Fig.~\ref{fig:emiss_cool}) and hotter (Fig.~\ref{fig:emiss_hot}) temperature intervals respectively.
The upper, middle and lower panels integrate over the $x$, $y$ and $z$ line of sights respectively. 
Contours of the $B_z$ field component at the photosphere have been over-plotted for the $z$ line of sight in order to visualise the photospheric field configuration.

The flux ropes of both eruptions can be seen in the cooler temperature interval (Fig.~\ref{fig:emiss_cool}a,b,c,d). In fact, these cool structures are the filament channels associated with the eruptions.
The second, and more energetic eruption appears as an 8-shape flux rope in the hotter temperature intervals (Fig.~\ref{fig:emiss_hot}d,e), having a hot core between $z=7.5-9.5$~Mm (Fig.~\ref{fig:emiss_hot}e).
The flux rope of the first eruption is not visible in the hotter temperature intervals, as the eruption is weaker. The hot structure in Fig.~\ref{fig:emiss_hot}b is the current sheet developing during the eruption.

An important difference between the two eruption is found at the location of the post-reconnection arcade. 
During both eruption, arcade-like loops form below the erupting flux rope (similar to the yellow lines in Fig.~\ref{fig:fieldlines2}e).
In the cooler temperature interval, the arcade is visible for both eruptions and is located below the lower current sheet. 
This is the bright band along the polarity inversion line, separating the inner polarities (e.g. Fig.~\ref{fig:emiss_cool}~c,f). 
However, in the hotter temperature interval, the arcade appears only in the second eruption (Fig.~\ref{fig:emiss_hot}). Therefore the arcade is flaring only in the second eruption and not the first. 


The reason for this is simply that the second eruption is more energetic. The flux rope associated with the second eruption accelerates more, becomes bigger in size and flux, and the eventual eruption has 6 times more kinetic energy. 
There are various reasons for this. During the eruptions, the magnetic energy (Fig.~\ref{fig:energy_time}) and photospheric unsigned flux is still increasing.
This suggests there is more magnetic free energy available for release come the second eruption. Also, since the magnetic lobes continue to expand with time, the topmost field strength actually decreases. Thus, reconnection at the middle current sheet will allow the flux rope to penetrate higher into the atmosphere (where the tension force is less) allowing the eruption to develop more before it eventually becomes suppressed. 
This further upwards rise allow more time for reconnection through the lower current sheet during the eruptive phase.
The extended duration of the reconnection at the lower current sheet during the eruptions is further affected by other parameters. The continued shearing and compression of the inner polarities, as also the rotation of the polarities, has brought them closer together by the onset of the second eruption. This causes field lines to interact with angles more favourable to efficient reconnection.  Also, the polarities of each of the bipoles have rotated more making the field lines more curved and thus increasing the angle of reconnection further. This results in faster jets and increased upwards tension from the U-shaped field lines. As discussed in \ref{subsec:second_eruption}, this will process assists the eruption extending the duration of the reconnection through the lower current sheet.
Similarly, this results in strong downwards release of tension from $\cap$-shaped lines that carry high temperature plasma to the post-reconnection arcade, but also compress the arcade, heating the plasma locally by shocks.

\section{Summary and Discussion}
\label{sec:discussion}

In this paper, we have presented results from a 3D flux emergence simulation of a quadrupolar region and analysed the subsequent eruptive dynamics. 
We placed a magnetic flux tube in the solar interior and triggered its emergence at two locations along its length to form a pair of bipoles at the photosphere. 
During the evolution of the system, the two initially separated bipoles converge, forming a strong $\delta$-shaped region between the two inner polarities of the quadrupolar region, as shown in Fig.~\ref{fig:magnetogram}. 
Inside the atmosphere, the field expands outwards forming two magnetic lobes that eventually interact through a series of current sheets.
At the PIL between the inner polarities, sheared field lines reconnect and form two flux ropes that erupt in a confined manner.

The two successive eruptions have distinct rise phases controlled by reconnection at the lower (below the flux rope), middle (above the flux rope) and upper current sheets (further above above the flux rope). 
In both cases, slow low-lying tether cutting reconnection through the lower current sheet gradually builds up a flux rope, which moves upwards due to magnetic pressure. 
At the same time, reconnection between the two lobes higher in the atmosphere (at the upper current sheet) forms field lines that retract down pushing against the flux rope, creating the middle current sheet between them.
The reconnection rate between the rising flux rope and the retracted field changes over time i) as the relative angle of the field lines of the two systems increase and ii) as the flux ropes moves upwards because of the emergence/shearing/rotation of the photospheric field.

The triggering of both eruptions occurs when the reconnection between the rising flux rope and the field directly above it becomes efficient. 
Our simulation assumes a non-magnetized atmosphere, therefore, there is no external reconnection between the quadrupolar system and any pre-existing atmospheric field. 
Despite the lack of any external field, the triggering mechanism of the eruptions is similar to the process of the ``external reconnection'', as both flux ropes erupt when the tension of the -dynamically formed- overlying field is reduced. It is well known that the relative angle between two magnetic field systems can produce or suppress an eruption \citep[e.g.][]{Archontis_etal2012}.
It is therefore very reasonable that in this work we find the triggering of both eruption to be affected again by the relative angle of the two interacting systems.

After the initial acceleration phase, both eruptions become confined by the overlying field (green and grey lines, Fig.~\ref{fig:fieldlines2}e) and the previously emerged stretched lines (pink lines, Fig.~\ref{fig:fieldlines2}e). The overlying strapping field is the result of the reconnection between the two magnetic lobes that initially emerged.
However, in principle, if the eruptions where stronger the strapping field could have been fully removed leading to an ejective eruption.
Therefore, our results indicates that depending on the internal structure of the field of a quadrupolar region, a flux rope can become eruptive even without any reconnection between the field of the quadrupolar region and an ambient external field. 

During the formation of the first flux rope, less magnetic energy is available. Also, the field that reconnects to form the flux rope is less sheared. This results to a smaller flux rope and an overall weaker eruption. This first eruptive flux rope almost fully reconnects with the field above and dissipates. This dissipation of the flux rope during the eruption is similar to the findings reported by \citet{Liu_etal2014c,Chintzoglou_etal2017}.
During the eruption, a cool erupting filament can be identified in emission proxy images, without any signature of flaring, as the eruption is short-lasting and weak.

The second eruption is more energetic than the first one. At later stages of the simulation, more flux has emerged and more energy is available to be released come the second eruption. 
In addition, the continued expansion of the lobes reduces the field strength and tension of the strapping field, allowing the flux rope to build up and rise more. 
However, the most important element, is that the low-lying field lines reconnecting to form the flux rope are more sheared, resulting to more effective reconnection. 
This results in a larger pre-eruptive flux rope. During the eruption, this results in powerful and longer lasting outflow from the current sheet below the flux rope, that assist the upwards acceleration of the rope. 
At the same time, the reconnection at the current sheet below the flux rope produces lines that add significant flux into the erupting flux rope, and increase its twist. 
This leads to the flux rope not being dissipated by the reconnection above it.
Instead, the flux rope's size is enhanced.
Also, the strong and longer lasting downflows from the current sheet below the flux rope result in the flaring of the post-reconnection arcade. 
The erupting flux rope can be identified in both hot and cooler temperatures as an 8-shaped flux rope in the corona.
The above process is an important result of our numerical investigation, as many observations suggest that a flux rope can form or be enhanced during confined flares \citep[e.g.][]{Guo_etal2012,Patsourakos_etal2013,Tziotziou_etal2013,Chintzoglou_etal2015,James_etal2017,James_etal2018,Liu_etal2018}. 
Our numerical model is in very good agreement with these observations, providing important insight on the confined-flare-to-flux-rope scenario.

Flux emergence simulations have studied quadrupolar configurations by assuming the emergence of a single $\Omega$-loop weakly twisted or non twisted \citep{Murray_etal2006,Archontis_etal2013,Syntelis_etal2015} flux tubes, the emergence of two $\Omega$-loop segments of a highly twisted flux tube \citep{Lee_etal2015,Fang_Fan_2015,Toriumi_etal2017b}, or the emergence of kink unstable flux tubes \citep[e.g.][]{Takasao_etal2015,Toriumi_etal2017b,Knizhnik_etal2018}. 
We will briefly discuss some differences between these approaches. \cite{Takasao_etal2015} noted that the kink unstable flux tubes and the two $\Omega$-loop  segments lead to some significant differences (see Introduction).
So, here, we will discuss the differences between the emergence of two $\Omega$-loop highly twisted flux tube segments against the emergence of a single $\Omega$-loop weakly twisted flux tube segment.
Both cases can lead to the simultaneous emergence of two bipoles of similar fluxes at the photopsphere, that are magnetically linked below the photosphere. Both cases eventually form a quadrupolar region with a strong PIL that is able to form a low-lying post-emergence flux rope \citep[e.g. comparison between][and our simulation]{Archontis_etal2013}. 

However, the two cases can potentially produce different dynamics. To examine these possibilities, we can make a comparison based on our knowledge of the emergence of a highly twisted single $\Omega$-loop. Such flux tubes can form a $\delta$-spot, flux ropes and eruptions \citep[e.g.][]{Leake_etal2014,Syntelis_etal2017,Syntelis_etal2019}, and can induce stronger rotation at the photospheric polarities \citep{Sturrock_etal2016}. 
Therefore, the emergence of two $\Omega$-loop segments of a highly twisted flux tube should induce more shear into the atmosphere by both the direct emergence of horizontal field and by the higher vorticity, in comparison to a single $\Omega$-loop weakly twisted flux tube of similar flux and radius. 
However, during the emergence of single $\Omega$-loop weakly twisted flux tube, the inner polarities spread more horizontally as they are less constrained by the azimuthal tension associated with the twist of the flux tube. Therefore, in this case, it is possible that the shear induced by horizontal shearing is more pronounced along a more elongated PIL.
Another difference between the two models could arise at the self-PILs of the two bipoles. 
A single $\Omega$-loop highly twisted flux tube emerge to form a bipolar region with a strong (self) PIL that is capable of building up new flux ropes \citep[e.g.][]{Syntelis_etal2017}. 
So, in principle, the emergence of two $\Omega$-loop segments of a highly twisted flux tube could form two filaments, one above each of the two self-PILs, similar to some observations \citep[e.g.][]{Chen_etal2018}. 
However, for our numerical setup and during the runtime of the simulation, we don't find such a case, and thus, further numerical investigation is required. It is possible that the emergence of two more localized $\Omega$-loop segments (i.e. smaller $\lambda$) is needed to produce strong self-PILs capable of building up flux ropes above them.
The self-PILs of the two bipoles in the cases of a single $\Omega$-loop weakly twisted flux tube are less strong and less shearing develops along them. So, flux ropes could be difficult, if not impossible, to form at these self-PILs, at least until the decay phase of the active region.

In the future, we aim to compare quadrupolar regions of different subphotospheric origin to study the in detail the differences in injection of free energy into the atmosphere, the location where it occurs and any subsequent eruptivity.

\begin{acknowledgements}
This project has received funding from the Science and Technology Facilities Council (UK) through the consolidated grant ST/S000402/1. 
The authors acknowledge support by the Royal Society grant RGF/EA/180232.
This work was supported by computational time granted from the Greek Research \& Technology Network (GRNET) in the National HPC facility - ARIS. The authors acknowledge support by the ERC synergy grant ``The Whole Sun''. 

\end{acknowledgements}

\bibliographystyle{aa}
\bibliography{references}

\end{document}